\documentclass[twocolumn]{aastex61}

\usepackage{times}
\usepackage{graphicx}
\usepackage{amssymb}
\usepackage{amsmath}
\usepackage{pifont}
\usepackage{graphics}
\usepackage{xcolor}
\usepackage{epsfig}
\usepackage{verbatim}
\usepackage{booktabs}

\newcommand{\bjdtdb}{BJD$_{\rm TDB}\ $}      
\newcommand{\feh}{\ensuremath{\left[{\rm Fe}/{\rm H}\right]}}  

\newcommand{\teff}{\ensuremath{T_{\rm eff}}}
\newcommand{\loggstar}{\ensuremath{\log{g_*}}}
\newcommand{\vsinistar}{\ensuremath{v\sin{I_*}}}
\newcommand{\msun}{\ensuremath{\,M_\Sun}}
\newcommand{\rsun}{\ensuremath{\,R_\Sun}}
\newcommand{\lsun}{\ensuremath{\,L_\Sun}}
\newcommand{\mj}{\ensuremath{\,M_{\rm J}}}
\newcommand{\rj}{\ensuremath{\,R_{\rm J}}}
\newcommand{\fave}{\langle F \rangle}
\newcommand{\fluxcgs}{\ensuremath{\rm 10^9 erg~s^{-1} cm^{-2}}}

\newcommand{\kms}{\ensuremath{\rm km\ s^{-1}}}

\newcommand{\mstar}{\ensuremath{M_{*}}}
\newcommand{\rstar}{\ensuremath{R_{*}}}

\begin{document}
\title{KELT-21\MakeLowercase{b}: A Hot Jupiter Transiting the Rapidly-Rotating Metal-Poor Late-A Primary of a Likely Hierarchical Triple System
}

\author{Marshall C. Johnson} 
\affiliation{Department of Astronomy, The Ohio State University, 140 West 18$^{th}$ Ave., Columbus, OH 43210, USA} 

\author{Joseph E. Rodriguez} 
\affiliation{Harvard-Smithsonian Center for Astrophysics, 60 Garden St., Cambridge, MA 02138, USA}

\author{George Zhou} 
\affiliation{Harvard-Smithsonian Center for Astrophysics, 60 Garden St., Cambridge, MA 02138, USA}
\affiliation{Hubble Fellow}

\author{Erica J. Gonzales} 
\affiliation{Department of Astronomy and Astrophysics, University of California, Santa Cruz, 1156 High St., Santa Cruz, CA 95064, USA}
\affiliation{NSF Graduate Fellow}

\author{Phillip A. Cargile} 
\affiliation{Harvard-Smithsonian Center for Astrophysics, 60 Garden St., Cambridge, MA 02138, USA}

\author{Justin R. Crepp} 
\affiliation{Department of Physics, University of Notre Dame, 225 Nieuwland Science Hall, Notre Dame, IN 46556, USA}

\author{Kaloyan Penev} 
\affiliation{Department of Physics, University of Texas at Dallas, Richardson, TX 75080, USA}

\author{Keivan G. Stassun} 
\affiliation{Department   of   Physics   and   Astronomy,    Vanderbilt   University,
Nashville, TN 37235, USA}

\author{B. Scott Gaudi} 
\affiliation{Department of Astronomy, The Ohio State University, 140 West 18$^{th}$ Ave., Columbus, OH 43210, USA}

\author{Knicole D. Col\'on}
\affiliation{NASA Goddard Space Flight Center, Greenbelt, MD 20771, USA }

\author{Daniel J. Stevens} 
\affiliation{Department of Astronomy, The Ohio State University, 140 West 18$^{th}$ Ave., Columbus, OH 43210, USA}


\author{Klaus G. Strassmeier}
\affiliation{Leibniz-Institut f\"ur Astrophysik Potsdam (AIP), An der Sternwarte 16, D-14482 Potsdam, Germany}

\author{Ilya Ilyin}
\affiliation{Leibniz-Institut f\"ur Astrophysik Potsdam (AIP), An der Sternwarte 16, D-14482 Potsdam, Germany}


\author{Karen A. Collins}
\affiliation{Harvard-Smithsonian Center for Astrophysics, 60 Garden St., Cambridge, MA 02138, USA}

\author{John F. Kielkopf}
\affiliation{Department of Physics and Astronomy, University of Louisville, Louisville, KY 40292, USA}

\author{Thomas E. Oberst}
\affiliation{Department of Physics, Westminster College, New Wilmington, PA 16172, USA}

\author{Luke Maritch} 
\affiliation{Department of Physical Sciences, Kutztown University, Kutztown, PA 19530, USA}

\author{Phillip A. Reed}
\affiliation{Department of Physical Sciences, Kutztown University, Kutztown, PA 19530, USA}

\author{Joao Gregorio}
\affiliation{Atalaia Group \& CROW Observatory, Portalegre, Portugal}

\author{Valerio Bozza}
\affiliation{Dipartimento di Fisica ``E. R. Caianiello'', Universit\`a di Salerno, Via Giovanni Paolo II 132, 84084 Fisciano (SA), Italy}
\affiliation{Istituto Nazionale di Fisica Nucleare, Sezione di Napoli, 80126 Napoli, Italy}

\author{Sebastiano Calchi Novati}
\affiliation{Dipartimento di Fisica ``E. R. Caianiello'', Universit\`a di Salerno, Via Giovanni Paolo II 132, 84084 Fisciano (SA), Italy}
\affiliation{IPAC, Mail Code 100-22, Caltech, 1200 East California Boulevard, Pasadena, CA 91125, USA}

\author{Giuseppe D'Ago}
\affiliation{INAF-Osservatorio Astronomico di Capodimonte, Salita Moiariello 16, 80131, Napoli, Italy}

\author{Gaetano Scarpetta}
\affiliation{Dipartimento di Fisica ``E. R. Caianiello'', Universit\`a di Salerno, Via Giovanni Paolo II 132, 84084 Fisciano (SA), Italy}
\affiliation{Instituto Internazionale per gli Alti Studi Scientifici (IIASS), Via G. Pellegrino 19, 84019 Vietri sul Mare (SA), Italy}

\author{Roberto Zambelli}
\affiliation{Societ\`a Astronomica Lunae, Castelnuovo Magra 19030, Italy}


\author{David W. Latham}
\affiliation{Harvard-Smithsonian Center for Astrophysics, 60 Garden St., Cambridge, MA 02138, USA}

\author{Allyson Bieryla}
\affiliation{Harvard-Smithsonian Center for Astrophysics, 60 Garden St., Cambridge, MA 02138, USA}

\author{William D. Cochran}
\affiliation{McDonald Observatory, University of Texas at Austin, 2515 Speedway, Stop C1400, Austin, TX 78712, USA}

\author{Michael Endl}
\affiliation{McDonald Observatory, University of Texas at Austin, 2515 Speedway, Stop C1400, Austin, TX 78712, USA}


\author{Jamie Tayar}
\affiliation{Department of Astronomy, The Ohio State University, 140 West 18$^{th}$ Ave., Columbus, OH 43210, USA}

\author{Aldo Serenelli}
\affiliation{Institute of Space Sciences (ICE, CSIC), Campus UAB, Carrer de Can Magrans S/N, E-08193, Barcelona, Spain}
\affiliation{Institut d'Estudis Espacials de Catalunya (IEEC), Edifici Nexus, C/Gran Capita, 2-4, E-08034, Barcelona, Spain}

\author{Victor Silva Aguirre}
\affiliation{Stellar Astrophysics Centre, Department of Physics and Astronomy, Aarhus University, Ny Munkegade 120, DK-8000 Aarhus C, Denmark}


\author{Seth P. Clarke}
\affiliation{Department  of  Physics  and  Astronomy,  Brigham  Young  University,
Provo, UT 84602, USA}

\author{Maria Martinez}
\affiliation{Department  of  Physics  and  Astronomy,  Brigham  Young  University,
Provo, UT 84602, USA}
\affiliation{University of California, Merced, Merced, CA 95343 USA}

\author{Michelle Spencer}
\affiliation{Department  of  Physics  and  Astronomy,  Brigham  Young  University,
Provo, UT 84602, USA}

\author{Jason Trump}
\affiliation{Department  of  Physics  and  Astronomy,  Brigham  Young  University,
Provo, UT 84602, USA}

\author{Michael D. Joner}
\affiliation{Department  of  Physics  and  Astronomy,  Brigham  Young  University,
Provo, UT 84602, USA}

\author{Adam G. Bugg}
\affiliation{Department  of  Physics  and  Astronomy,  Brigham  Young  University,
Provo, UT 84602, USA}

\author{Eric G. Hintz}
\affiliation{Department  of  Physics  and  Astronomy,  Brigham  Young  University,
Provo, UT 84602, USA}

\author{Denise C. Stephens}
\affiliation{Department  of  Physics  and  Astronomy,  Brigham  Young  University,
Provo, UT 84602, USA}

\author{Anicia Arredondo}
\affiliation{Department of Astronomy, Wellesley College, Wellesley, MA 02481, USA}
\affiliation{University of Central Florida, Orlando, FL 32816 USA}

\author{Anissa Benzaid}
\affiliation{Department of Astronomy, Wellesley College, Wellesley, MA 02481, USA}
\affiliation{Boston University, Boston, MA 02215, USA}

\author{Sormeh Yazdi}
\affiliation{Department of Astronomy, Wellesley College, Wellesley, MA 02481, USA}
\affiliation{Massachusetts Institute of Technology, Cambridge, MA 02139, USA}

\author{Kim K. McLeod}
\affiliation{Department of Astronomy, Wellesley College, Wellesley, MA 02481, USA}

\author{Eric L. N. Jensen}
\affiliation{Department of Physics and Astronomy, Swarthmore College, Swarthmore, PA 19081, USA}

\author{Daniel A. Hancock}
\affiliation{Department of Physics and Astronomy, University of Wyoming, Laramie, WY 82071, USA}

\author{Rebecca L. Sorber}
\affiliation{Department of Physics and Astronomy, University of Wyoming, Laramie, WY 82071, USA}

\author{David H. Kasper}
\affiliation{Department of Physics and Astronomy, University of Wyoming, Laramie, WY 82071, USA}

\author{Hannah Jang-Condell}
\affiliation{Department of Physics and Astronomy, University of Wyoming, Laramie, WY 82071, USA}


\author{Thomas G. Beatty}
\affiliation{Department of Astronomy \& Astrophysics, The Pennsylvania State University, 525 Davey Lab, University Park, PA 16802, USA }
\affiliation{Center for Exoplanets and Habitable Worlds, The Pennsylvania State University, 525 Davey Lab, University Park, PA 16802, USA }

\author{Thorsten Carroll}
\affiliation{Leibniz-Institut f\"ur Astrophysik Potsdam (AIP), An der Sternwarte 16, D-14482 Potsdam, Germany}

\author{Jason Eastman}
\affiliation{Harvard-Smithsonian Center for Astrophysics, 60 Garden St., Cambridge, MA 02138, USA}

\author{David James}
\affiliation{Harvard-Smithsonian Center for Astrophysics, 60 Garden St., Cambridge, MA 02138, USA}

\author{Rudolf B. Kuhn}
\affiliation{South African Astronomical Observatory, PO Box 9, Observatory, 7935 Cape Town, South Africa}
\affiliation{Southern African Large Telescope, PO Box 9, Observatory, 7935 Cape Town, South Africa}

\author{Jonathan Labadie-Bartz}
\affiliation{Department of Physics, Lehigh University, 16 Memorial Drive East, Bethlehem, PA 18015, USA}

\author{Michael B. Lund}
\affiliation{Department of Physics and Astronomy, Vanderbilt University, Nashville, TN 37235, USA}

\author{Matthias Mallonn}
\affiliation{Leibniz-Institut f\"ur Astrophysik Potsdam (AIP), An der Sternwarte 16, D-14482 Potsdam, Germany}

\author{Joshua Pepper}
\affiliation{Department of Physics, Lehigh University, 16 Memorial Drive East, Bethlehem, PA 18015, USA}

\author{Robert J. Siverd}
\affiliation{Las Cumbres Observatory, 6740 Cortona Dr., Ste 102, Goleta, CA 93117, USA}

\author{Xinyu Yao}
\affiliation{Department of Physics, Lehigh University, 16 Memorial Drive East, Bethlehem, PA 18015, USA}


\author{David H. Cohen}
\affiliation{Department of Physics and Astronomy, Swarthmore College, Swarthmore, PA 19081, USA}

\author{Ivan A. Curtis}
\affiliation{ICO, Adelaide, South Australia, Australia}

\author{D. L. DePoy}
\affiliation{George P. and Cynthia Woods Mitchell Institute for Fundamental Physics and Astronomy, Texas A\&M University, College Station, TX 77843, USA}
\affiliation{Department of Physics and Astronomy, Texas A\&M University, College Station, TX 77843, USA}

\author{Benjamin J. Fulton}
\affiliation{California Institute of Technology, Pasadena, CA 91125, USA}
\affiliation{Texaco Fellow}

\author{Matthew T. Penny} 
\affiliation{Department of Astronomy, The Ohio State University, 140 West 18$^{th}$ Ave., Columbus, OH 43210, USA}

\author{Howard Relles}
\affiliation{Harvard-Smithsonian Center for Astrophysics, 60 Garden St., Cambridge, MA 02138, USA}

\author{Christopher Stockdale}
\affiliation{Hazelwood Observatory, Churchill, Victoria, Australia}

\author{Thiam-Guan Tan}
\affiliation{Perth Exoplanet Survey Telescope, Perth, Australia}

\author{Steven Villanueva, Jr.} 
\affiliation{Department of Astronomy, The Ohio State University, 140 West 18$^{th}$ Ave., Columbus, OH 43210, USA}
\affiliation{NSF Graduate Fellow}

\shorttitle{KELT-21\MakeLowercase{b}}

\begin{abstract}

We present the discovery of KELT-21b, a hot Jupiter transiting the $V=10.5$ A8V star HD 332124. The planet has an orbital period of $P=3.6127647\pm0.0000033$ days and a radius of $1.586_{-0.040}^{+0.039}$ \rj. We set an upper limit on the planetary mass of $M_P<3.91$ \mj\ at $3\sigma$ confidence. We confirmed the planetary nature of the transiting companion using this mass limit and Doppler tomographic observations to verify that the companion transits HD 332124. These data also demonstrate that the planetary orbit is well-aligned with the stellar spin, with a sky-projected spin-orbit misalignment of $\lambda=-5.6_{-1.9}^{+1.7 \circ}$.
The star has $\teff=7598_{-84}^{+81}$ K, $M_*=1.458_{-0.028}^{+0.029} \msun$, $R_*=1.638\pm0.034 \rsun$, and $\vsinistar=146$ km s$^{-1}$, the highest projected rotation velocity of any star known to host a transiting hot Jupiter.
The star also appears to be somewhat metal-poor and $\alpha$-enhanced, with $\feh=-0.405_{-0.033}^{+0.032}$ and [$\alpha$/Fe]$=0.145 \pm 0.053$; these abundances are unusual, but not extraordinary, for a young star with thin-disk kinematics like KELT-21.
High-resolution imaging observations revealed the presence of a pair of stellar companions to KELT-21, located at a separation of 1\farcs2 and with a combined contrast of $\Delta K_S=6.39 \pm 0.06$ with respect to the primary. Although these companions are most likely physically associated with KELT-21, we cannot confirm this with our current data. If associated, the candidate companions KELT-21 B and C would each have masses of $\sim0.12$ \msun, a projected mutual separation of $\sim20$ AU, and a projected separation of $\sim500$ AU from KELT-21. KELT-21b may be one of only a handful of known transiting planets in hierarchical triple stellar systems.

\end{abstract}


\section{Introduction}
\label{sec:Intro}

Most of the currently known exoplanets orbit relatively cool, low-mass (FGKM) stars, with $\teff<6500$ K. The reason for this is likely observational bias rather than an actual paucity of planets around hotter stars: traditionally, precise radial velocity observations have been the dominant method to find exoplanets, and, more recently, confirm transiting giant planets. Hot stars, however, typically rotate rapidly. Above the Kraft break \citep{Kraft:1967}, at $\teff\sim6250$~K, stars no longer have the thick surface convective zones necessary to maintain a strong magnetic dynamo which can efficiently transport angular momentum to the outgoing stellar wind. More massive stars therefore tend to retain their initial rapid rotation throughout their main sequence lifetimes. Typical $\vsinistar$ values are in excess of 100 \kms\ for the entire B9-F2 spectral type range \citep{Royer:2007}. The rotational broadening of these stars' lines and the paucity of absorption lines from their hot atmospheres makes it difficult or impossible to measure the stellar reflex motion due to even giant planets. For this reason, stars hotter than mid-F have typically been ignored by planet surveys. Planets around these stars were too difficult to find with radial velocities or confirm as transiting planets, and massive stars are too rare to cause a significant number of microlensing events.

There are, however, other ways to discover planets around early-type stars. Direct imaging observations are largely insensitive to the stellar properties. Indeed, early-type stars are {\it more} amenable to direct imaging observations than are solar-type stars because they are on average younger and their planets therefore hotter and brighter. Many of the known directly-imaged planets are around such stars \citep[e.g.,][]{Marois:2008,Marois:2010,Lagrange:2010}. 

Planets can also be found around these stars by observing them after they have left the main sequence. As they evolve the stars cool and expand, slowing their rotation and increasing the number of spectral lines, thus becoming amenable to precise radial velocity observations. Radial velocity surveys have discovered a large number of giant planets around intermediate-mass subgiant and giant stars \citep[e.g.,][]{Johnson:2011,Reffert:2015}, which suggest that the frequency of giant planets around A stars may be higher than that around FGK stars. The issue of whether these ``retired A stars'' are actually intermediate-mass stars, or rather lower-mass interlopers, is, however, controversial \citep[e.g.,][]{Lloyd:2011,Schlaufman:2013,Johnson:2013,Stello:2017}.

Despite the success of direct imaging and radial velocity surveys, there are still limitations to these methods as far as probing the overall planetary populations of A and early F stars. Principally, neither method can probe the close-in, short-period population of planets, similar to those that {\it Kepler} has discovered around lower-mass stars. The angular separations between these planets and their host stars are much too small to be resolved with current or future direct imaging facilities, and such short-period planets will have been engulfed and destroyed as the stellar radii expand during the evolution off of the main sequence. 

Although early-type stars have typically been ignored by transit surveys, efforts to discover planets around these stars have been increasingly successful. The first transiting planet to be discovered around a hot, rapidly rotating star was WASP-33b \citep{CollierCameron2010}. It was confirmed using a combination of Doppler tomography (where the perturbation to the rotationally broadened stellar line profile during the transit due to the Rossiter-McLaughlin effect is spectroscopically resolved, showing that the planet candidate orbits the target star and is not a background eclipsing binary), and low-precision radial velocity observations (showing that the transiting object has a mass below the substellar limit). Since then, a growing number of transiting hot Jupiters around rapidly rotating A and early F stars have been discovered and (in most cases) confirmed using Doppler tomography, such as CoRoT-11b \citep{Gandolfi:2010,Gandolfi:2012}, Kepler-13Ab \citep{Szabo:2011,Johnson:2014}, HAT-P-57b \citep{Hartman:2015}, MASCARA-1b \citep{Talens:2017b}, and XO-6b \citep{Crouzet:2017}.

The Kilodegree Extremely Little Telescope \citep[KELT;][]{Pepper:2003,Pepper:2007,Pepper:2012} project has been particularly effective in finding hot Jupiters around A and early F stars. Indeed, more than half of the KELT planets discovered to date orbit stars above the Kraft break \citep[][]{Siverd:2012,Pepper:2013,Collins:2014,Bieryla:2015,Stevens:2016,Zhou:2017,McLeod:2017,Temple:2017,Gaudi:2017,Lund:2017,Siverd:2017}, or, in the case of KELT-11, was above the Kraft break when it was on the main sequence \citep{Pepper:2016}. As discussed in \cite{Lund:2017}, this is partially by design and Malmquist bias, and partially by accident. As KELT was designed to target brighter stars ($8<V<11$) than other transit surveys, hot, bright stars are overrepresented in its sample due to the larger volume probed. Additionally, many of the transiting hot Jupiters around cooler stars in this magnitude range had already been discovered by the time KELT had obtained a sufficient quantity of data to find planets. Together, these two forces drove us towards the discovery of planets around hot stars.

Another region of parameter space that has not been fully explored is hot Jupiter occurrence as a function of stellar multiplicity. While there are a large number of known hot Jupiters in binary stellar systems--indeed, \cite{Ngo:2016} found a higher occurrence rate of stellar binary companions for hot Jupiter hosts than for field stars--only a handful of hot Jupiters are known in higher-order stellar systems. Four transiting hot Jupiters are known in hierarchical triple systems: KELT-4 \citep{Eastman:2016}, HAT-P-8, WASP-12 \citep{Bechter:2014}, and Kepler-13 \citep{Santerne:2012}, and none in higher-order multiples. In the first three of these the hot Jupiter orbits the primary star, the mass ratios between the primary and secondary/tertiary stars are large, and the companions were resolved with direct imaging. In Kepler-13, the primary and secondary stars are of a similar mass, while the tertiary star is much less massive, and the tertiary was discovered through radial velocity observations of the secondary \citep{Santerne:2012,Johnson:2014}. Transit survey follow-up is typically biased against the latter type of system; visual binaries are often excluded from survey target lists, and the presence of multiple lines in a spectrum (especially if some of them are moving) is typically taken as reason to conclude that a planet candidate is a false positive, even if such a system could in fact host a planet \citep[see][for a recent case of a confirmed planet discovered in a binary stellar system with two sets of lines in the spectra]{Siverd:2017}. 
Planets in hierarchical triple systems are also of interest because their dynamics are richer than those of hot Jupiters in stellar binaries \citep[e.g.,][]{Hamers:2017,Fang:2017}; certain configurations could enhance the efficiency of hot Jupiter formation by high-eccentricity (Kozai-Lidov) migration \citep{Hamers:2017}. Such extreme systems help us test the limits of planet formation and migration.

Here we present the discovery of a new transiting hot Jupiter, KELT-21b, confirmed using Doppler tomography. KELT-21 is not only the most rapidly rotating star known to host a transiting giant planet (with $\vsinistar=146$ \kms), but also is likely one of the few planet host stars known to be part of a hierarchical triple system.

\section{Discovery and Follow-Up Observations}
\label{sec:Obs}

\subsection{Discovery}
\label{sec:Discovery}

\begin{figure}
\centering 
\includegraphics[width=1.1\columnwidth, angle=0, trim = 0 5.2in 0 2.6in]{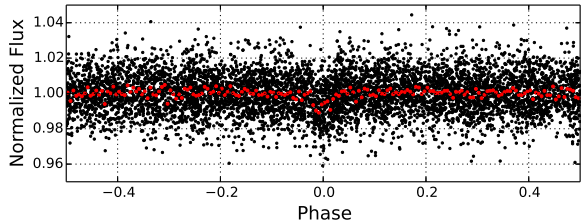}
\caption{\footnotesize Discovery light curve for KELT-21b based on 6783 observations from the KELT-North telescope.  The data have been phase-folded on the preliminary value for the period, 3.612782 days. The black points depict the data, while the red points are the data binned on a 5-minute time scale.}
\label{fig:DiscoveryLC}
\end{figure}

The star HD 332124 was observed by the KELT-North telescope in KELT North Field 11 (KN11; center coordinates RA=$19^h27^m00\fs0$, Dec=+31\arcdeg39\arcmin56\farcs2 J2000.0). This is the same field that also yielded our other two most rapidly rotating planet hosts, KELT-9b \citep{Gaudi:2017} and KELT-20b/MASCARA-2b \citep{Lund:2017,Talens:2017}, as well as KELT-8b \citep{Fulton:2015}. In the analysis of 6783 observations of HD 332124 obtained between 2007 May 29 and 2014 Nov 25 UT, we identified a candidate transit signal with a period of 3.612782 days and a depth of $\sim1\%$; the target was given the KELT candidate ID of KC11C039077. Our data reduction and analysis and candidate selection procedures are described in \cite{Siverd:2012}. We show the KELT photometry of KELT-21 in Fig.~\ref{fig:DiscoveryLC}, and list the literature parameters of the system in Table~\ref{tab:LitProps}.

\begin{table}
\footnotesize
\centering
\caption{System Properties of KELT-21 }
\begin{tabular}{llcc}
  \hline
  \hline
Other identifiers\dotfill & 
        \multicolumn{3}{l}{HD 332124} \\
      & \multicolumn{3}{l}{TYC 2676-1274-1}				\\
	  & \multicolumn{3}{l}{2MASS J20191200+3234517}  \\
	  & \multicolumn{3}{l}{TIC 203189770}
\\
\hline
Parameter & Description & Value & Ref. \\
\hline
$\alpha_{\rm J2000}$\dotfill	&Right Ascension (RA)\dotfill & $20^h19^m12\fs004$			& 1	\\
$\delta_{\rm J2000}$\dotfill	&Declination (Dec)\dotfill & +32\arcdeg34\arcmin51\farcs77			& 1	\\
\\
$B_{\rm T}$\dotfill			&Tycho $B_{\rm T}$ mag.\dotfill & 10.76 $\pm$ 0.04		& 1	\\
$V_{\rm T}$\dotfill			&Tycho $V_{\rm T}$ mag.\dotfill & 10.48 $\pm$ 0.04		& 1	\\

\\
$J$\dotfill			& 2MASS $J$ mag.\dotfill & 10.149  $\pm$ 0.022		& 2	\\
$H$\dotfill			& 2MASS $H$ mag.\dotfill & 10.121 $\pm$ 0.021	& 2	\\
$K_{\rm S}$\dotfill			& 2MASS $K_{\rm S}$ mag.\dotfill & 10.090 $\pm$ 0.014	& 2	\\
\\
\textit{W1}\dotfill		& \textit{WISE1} mag.\dotfill & $10.064 \pm	0.022$		& 3	\\
\textit{W2}\dotfill		& \textit{WISE2} mag.\dotfill & $10.106 \pm	0.021$		& 3 \\
\textit{W3}\dotfill		& \textit{WISE3} mag.\dotfill & $10.252 \pm	0.069$		& 3	\\
\textit{W4}\dotfill		& \textit{WISE4} mag.\dotfill & $>9.158$		& 3	\\
\\
$\mu_{\alpha}$\dotfill		& Gaia DR1 proper motion\dotfill & 1.933 $\pm$ 0.791 		& 4 \\
                    & \hspace{3pt} in RA (mas yr$^{-1}$)	& & \\
$\mu_{\delta}$\dotfill		& Gaia DR1 proper motion\dotfill 	&  -1.317 $\pm$ 0.804 &  4 \\
                    & \hspace{3pt} in DEC (mas yr$^{-1}$) & & \\
\\
$RV^{*}$\dotfill & Systemic radial \hspace{9pt}\dotfill  & $-13.0 \pm 1.0$ & \S\ref{sec:Spectra} \\
     & \hspace{3pt} velocity (\kms)  & & \\
$\vsinistar$\dotfill &  Stellar rotational \hspace{7pt}\dotfill &  $146.03 \pm 0.48$ & \S\ref{sec:GlobalFit} \\
                 & \hspace{3pt} velocity (\kms)  & & \\
Spec. Type\dotfill & Spectral Type\dotfill & A8V & \S\ref{sec:GlobalFit} \\
Age\dotfill & Age (Gyr)\dotfill & $1.6 \pm 0.1$ & \S\ref{sec:Evol} \\
$d$\dotfill & Distance (pc)\dotfill & $415 \pm 49$ & 4 \\
$\Pi$\dotfill & Parallax (mas) \dotfill & $2.41 \pm 0.28$ & 4 \\
$A_V$\dotfill & Visual extinction (mag) & $0.00_{-0.00}^{+0.34}$ & \S\ref{sec:SED} \\
$U^{\dagger}$\dotfill & Space motion (\kms)\dotfill & $9.2 \pm 1.9$  & \S\ref{sec:UVW} \\
$V$\dotfill       & Space motion (\kms)\dotfill & $-0.5 \pm 1.1$ & \S\ref{sec:UVW} \\
$W$\dotfill       & Space motion (\kms)\dotfill & $8.9 \pm 1.9$ & \S\ref{sec:UVW} \\
\hline
\hline
\end{tabular}
\begin{flushleft} 
 \footnotesize{ \textbf{\textsc{NOTES:}}
 $^{*}RV$ is defined on the IAU system.
 
$^{\dagger}U$ is positive in the direction of the Galactic Center. 
    References are: $^1$\citet{Hog:2000}, $^2$\citet{Cutri:2003}, $^3$\citet{Cutri:2014},$^4$\citet{Brown:2016} Gaia DR1 (http://gea.esac.esa.int/archive/), corrected for the position-dependent systematic offset found by \cite{Stassun:2016}.
    
}
\end{flushleft}
\label{tab:LitProps}
\end{table}

\subsection{Photometric Follow-up from KELT-FUN}
\label{sec:Photom}

After identification of the candidate transit signal in the KELT-North photometry, we obtained follow-up transit photometry of this target using the KELT Follow-Up Network (KFUN; Collins et al.\ in prep). The purpose of these observations was to first confirm the existence of a transit event for this star, and then check that the transit shape was consistent with that of a planet and that the transit was achromatic. We describe the facilities and observations obtained briefly in the following sections; KELT-21 passed all of these tests, and so we then pursued spectroscopic observations to confirm the planetary nature of the candidate (\S\ref{sec:Spectra}). We scheduled observations using a modified version of the TAPIR observation planning software \citep{Jensen:2013}, and all observations were reduced using the AstroImageJ package\footnote{http://www.astro.louisville.edu/software/astroimagej/} \citep{Collins:2013,Collins:2017}. 

\begin{figure}
\vspace{.0in}
\includegraphics[width=1\linewidth,height=5in]{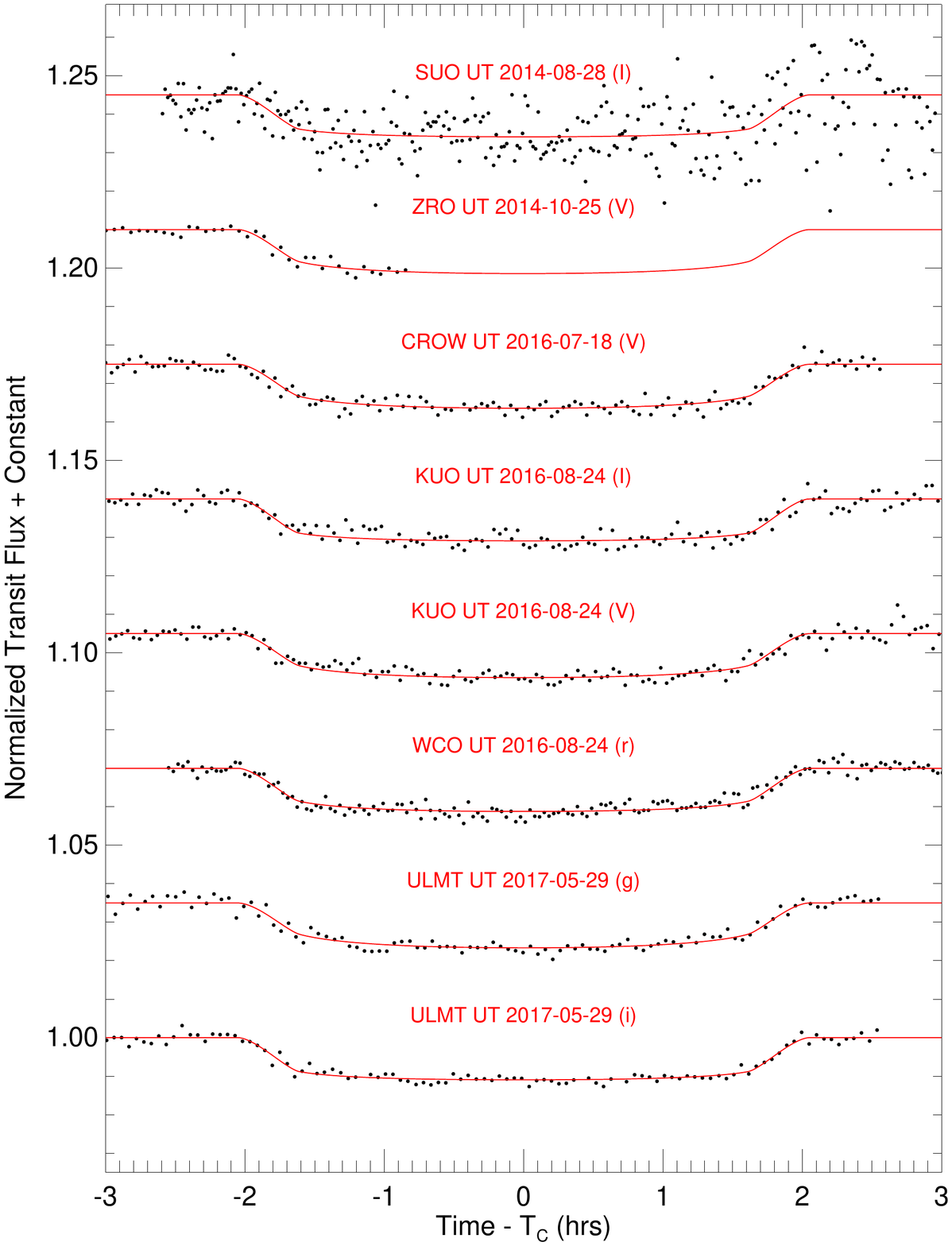}
\vspace{-.25in}
\includegraphics[width=1\linewidth, trim = 0 0.9in 0 5.5in]{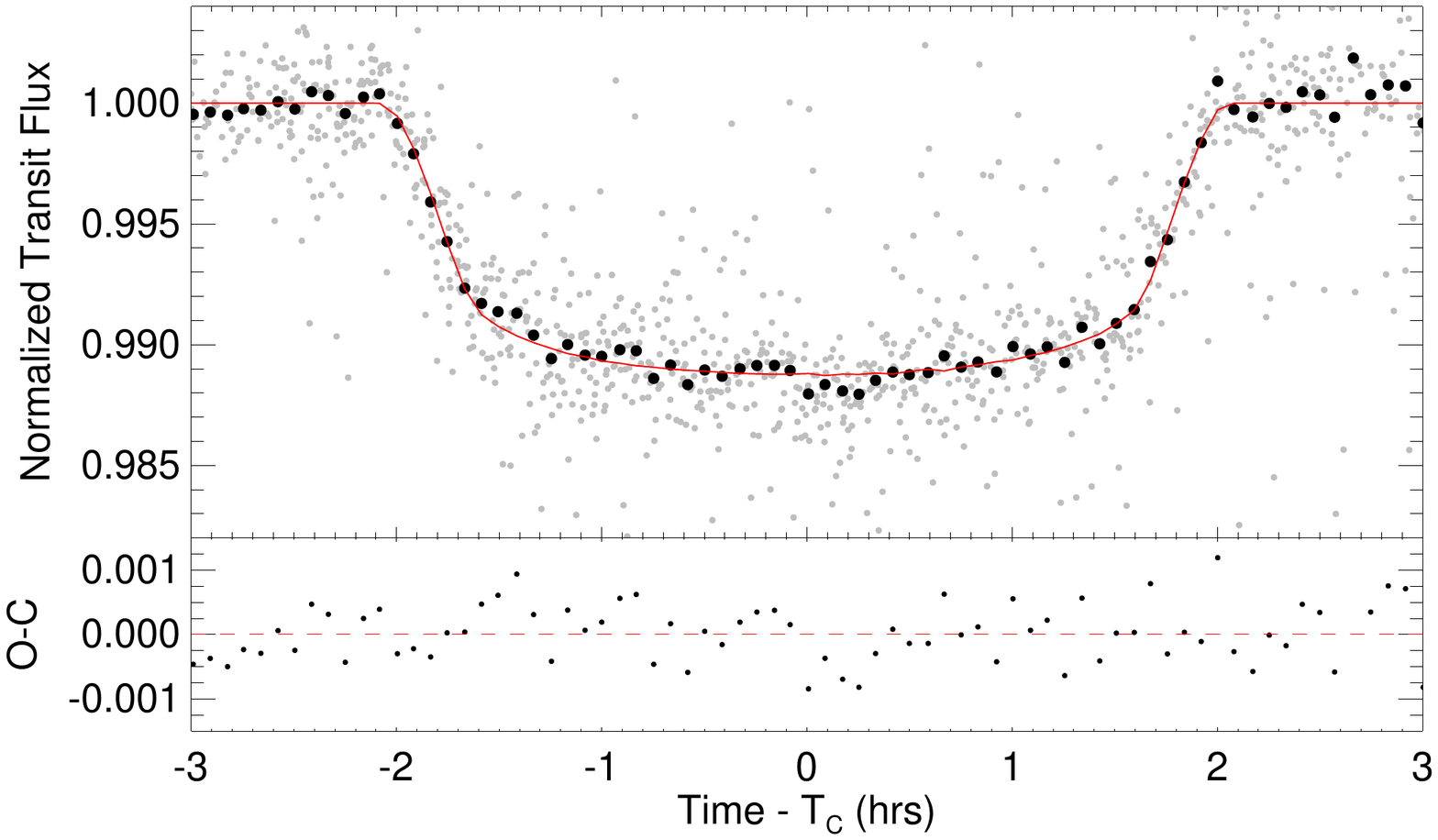}
\caption{(Top) Transit light curves of KELT-21b from the KELT Follow-Up Network. The red line represents the best fit model from the global fit in that photometric band. Each light curve is offset vertically by an arbitrary amount for clarity. (Bottom) All follow-up transits combined into one light curve (grey) and a 5 minute binned light curve (black). The red line is the combined and binned models for each transit. We emphasize that the combined light curve is only for display purposes; the individual transit light curves were used in our analysis.  
}
\label{fig:All_light curve} 
\end{figure}

We observed portions of seven transits of KELT-21b between 2014 Aug 7 and 2017 May 29 UT; two of these were observed simultaneously in two different filters, for a total of nine transit light curves. 
Details of these are listed in Table~\ref{tab:Photom}, and the data are shown in Fig.~\ref{fig:All_light curve}.

\subsubsection{Salerno University Observatory}

We observed the transit of 2014 Aug 28 at the Salerno University Observatory, located in Fisciano, Italy. 
At the time of the observations it hosted a Celestron C14 telescope (0.35~m aperture) on a German equatorial mount, with an SBIG ST2000XM CCD and standard Bessell filters; the observations were conducted in the $I$ band. The field of view was 11'x14' with a plate scale of 0\farcs59 pix$^{-1}$.

\subsubsection{Roberto Zambelli's Observatory}

We observed the ingress of the transit of 2014 Oct 25 at Zambelli's Robotic Observatory (ZRO) in Sarzana, Italy. We used a Meade 12'' (0.3048 m) f/10 telescope with a focal reducer giving a final value of f/6.3. The images were captured using an SBIG ST8XME CCD. The CCD has $1530\times1020$ pixels and a $23.'52\times15.'68$ field of view, giving a plate scale of 0\farcs92 pix$^{-1}$. The observations were obtained in the $V$ band with an exposure length of 200 seconds. Strong winds halted the observations about an hour before mid-transit.

\subsubsection{Canela's Robotic Observatory}

We observed a transit of KELT-21b on 2016 Jul 18 from Canela's Robotic Observatory (CROW) in Portalegre, Portugal. Located at an altitude of 600 m, the observatory is equipped with a Meade SCT 30 cm telescope with f/5.56 and an SBIG ST-10XME CCD camera with a KAF3200ME detector, giving a plate scale of 0\farcs82 pix$^{-1}$. We observed with a Johnson $V$ filter manufactured by Custom Scientific.

\subsubsection{Kutztown University Observatory}

We observed KELT-21 from the Kutztown University Observatory (KUO) for seven consecutive hours on 2016 Aug 24 in alternating $V$ and $I$ filters, covering the full $\sim4$-hour transit of KELT-21b and an additional 3 hours of pre-ingress and post-egress baseline.  A total of 332 images were collected (166 in each band) with exposure times of 60 s each.  We used KUO's 0.6 m f/8 Ritchey-Chr\'{e}tien optical telescope and an SBIG STXL-6303E CCD camera.  The detector's array of $3072\times2048$ 9\,$\mu$m pixels provides a $19.'5\times13.'0$ field of view, and with $2\times2$ binning, the effective plate scale is $0\farcs76$ pix$^{-1}$.  The CCD was kept at an operating temperature of $-25^{\circ}$ C.  KUO is located on the campus of Kutztown University in Kutztown, Pennsylvania, USA.

\subsubsection{Westminster College Observatory}

We observed a full transit of KELT-21b from the Westminster College Observatory (WCO), Pennsylvania, USA, on 2016 Aug 24 UT in the $r'$ filter. The observations employed a 0.35 m $f/11$ Celestron C14 Schmidt-Cassegrain telescope and an SBIG STL-6303E CCD camera with a $\sim$ 3k $\times$ 2k array of 9 $\mu$m pixels, yielding a $24\arcmin \times 16\arcmin$ field of view and $1\farcs44$ pix$^{-1}$ plate scale at $3 \times 3$ pixel binning.

\subsubsection{University of Louisville Manner Telescope}

We observed a full transit of KELT-21b using the University of Louisville Manner Telescope (ULMT) located at the Mt.\ Lemmon summit of Steward Observatory, Arizona, USA, on 2017 May 29 UT. Exposures were taken in alternating $g$ and $i$ filters yielding pseudo-simultaneous observations in the two bands. The observations employed a 0.6 m f/8 RC Optical Systems Ritchey-Chr\'{e}tien telescope and an SBIG STX-16803 CCD camera with a 4k$\times$4k array of 9\,$\mu$m pixels, yielding a $26\farcm6 \times 26\farcm6$ field of view and $0\farcs39$ pix$^{-1}$ plate scale. These photometric observations were obtained simultaneously with the spectroscopic observations with LBT/PEPSI and TRES described below.

\begin{table*}
 \footnotesize
 \centering
 \setlength\tabcolsep{1.5pt}
 \caption{Photometric follow-up observations of KELT-21\MakeLowercase{b}}
 \begin{tabular}{lcccccccc}
   \hline
   \hline

Observatory & Location & Aperture & Plate scale& Date      & Filter  & Exposure & Detrending parameters$^a$  \\
            &          & (m)      & ($\rm \arcsec~pix^{-1}$)& (UT) &        &  Time (s) &  & \\
\hline

SUO & Fisciano Salerno, Italy & 0.3556 & 0.59 & 2014 Aug 28 & $I$ & 60 & Airmass \\

ZRO & Sarzana, Italy & 0.3048 & 0.92 & 2014 Oct 25 & $V$ & 200 & Airmass\\ 
CROW & Portalegre, Portugal & 0.3048 & 0.82 & 2016 Jul 18 & $V$ & 120 & Airmass, Meridian Flip\\ 
KUO & PA, USA & 0.6096 & 0.76 & 2016 Aug 24 & $V$ & 60 & Airmass\\ 
KUO & PA, USA & 0.6096 & 0.76 & 2016 Aug 24 & $I$ & 60 & Airmass\\ 
WCO & PA, USA & 0.35 & 1.44 & 2016 Aug 24 & $r'$ & 15 & Airmass\\ 
ULMT & AZ, USA & 0.6096 & 0.39 & 2017 May 29 & $g'$ & 50 & Airmass\\ 
ULMT & AZ, USA & 0.6096 & 0.39 & 2017 May 29 & $i'$ & 100 & Airmass\\ 

\hline
\hline

\end{tabular}
\begin{flushleft}
  \footnotesize{ \textbf{\textsc{NOTES:}}  $^a$Photometric parameters allowed to vary in global fits and described in the text. Abbreviations: SUO: Salerno University Observatory; 
  ZRO: Zambelli's Robotic Observatory; CROW: Canela's Robotic Observatory; KUO: Kutztown Observatory, Kutztown University; WCO: Westminster College Observatory; ULMT: University of Louisville-Manner Telescope.} 
\end{flushleft}
\label{tab:Photom}
\end{table*}

\subsection{Spectroscopic Follow-up}
\label{sec:Spectra}

As the photometric follow-up campaign confirmed the existence of transits for KELT-21 and showed that the transits were both achromatic and with a shape appropriate for a transiting planet, we began spectroscopic follow-up observations. These consisted of three steps: first, reconnaissance spectroscopy to measure the stellar parameters; second, low-precision radial velocity (RV) observations to exclude large radial velocity variations that would have indicated that the transiting object was an M dwarf or a brown dwarf; and, finally, Doppler tomographic observations to confirm that the transiting object transits the star KELT-21. In the following sections we describe these observations.

\begin{figure}
\includegraphics[width=1\linewidth, trim = 0 1.5in 0 5.5in]{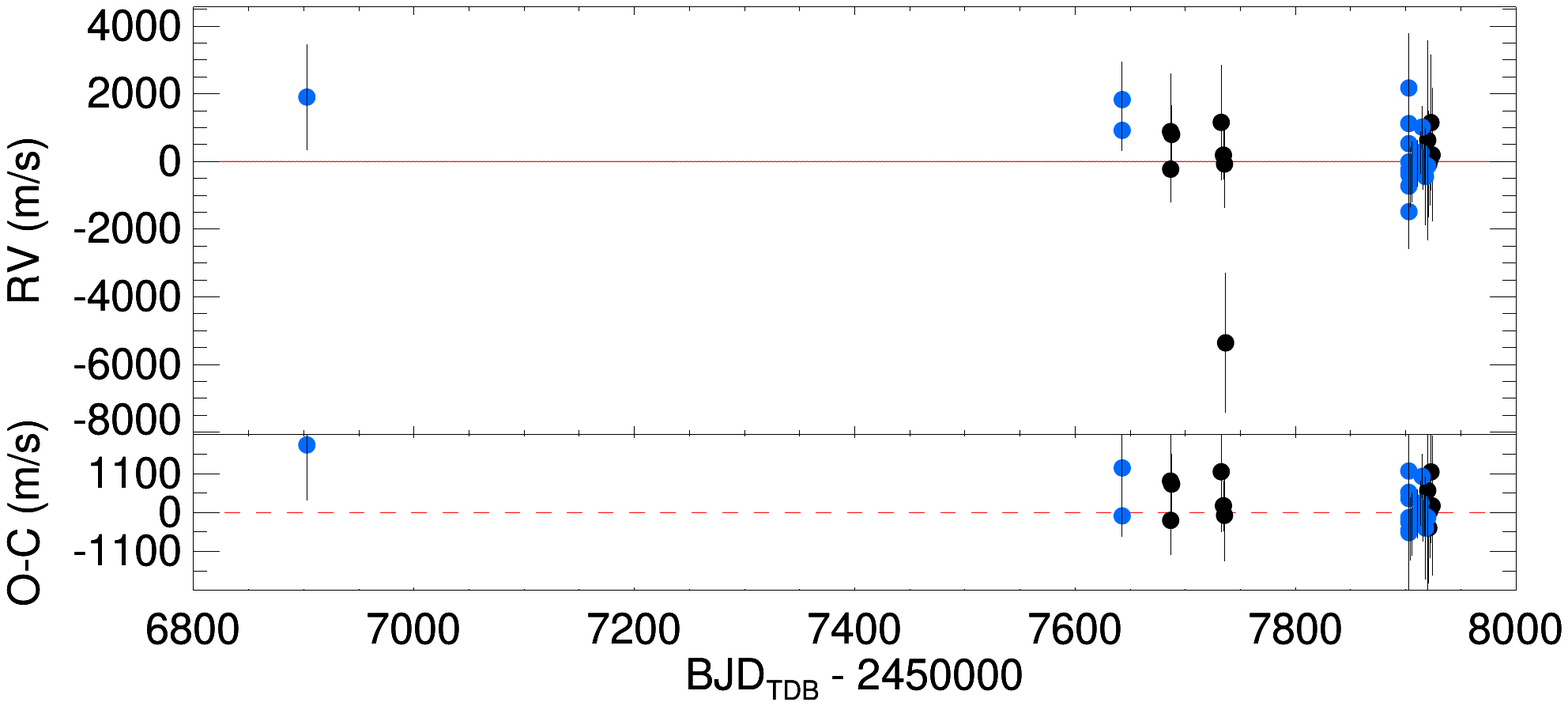}
   \vspace{-.3in}
\includegraphics[width=1\linewidth, trim = 0 1.5in 0 6.0in]{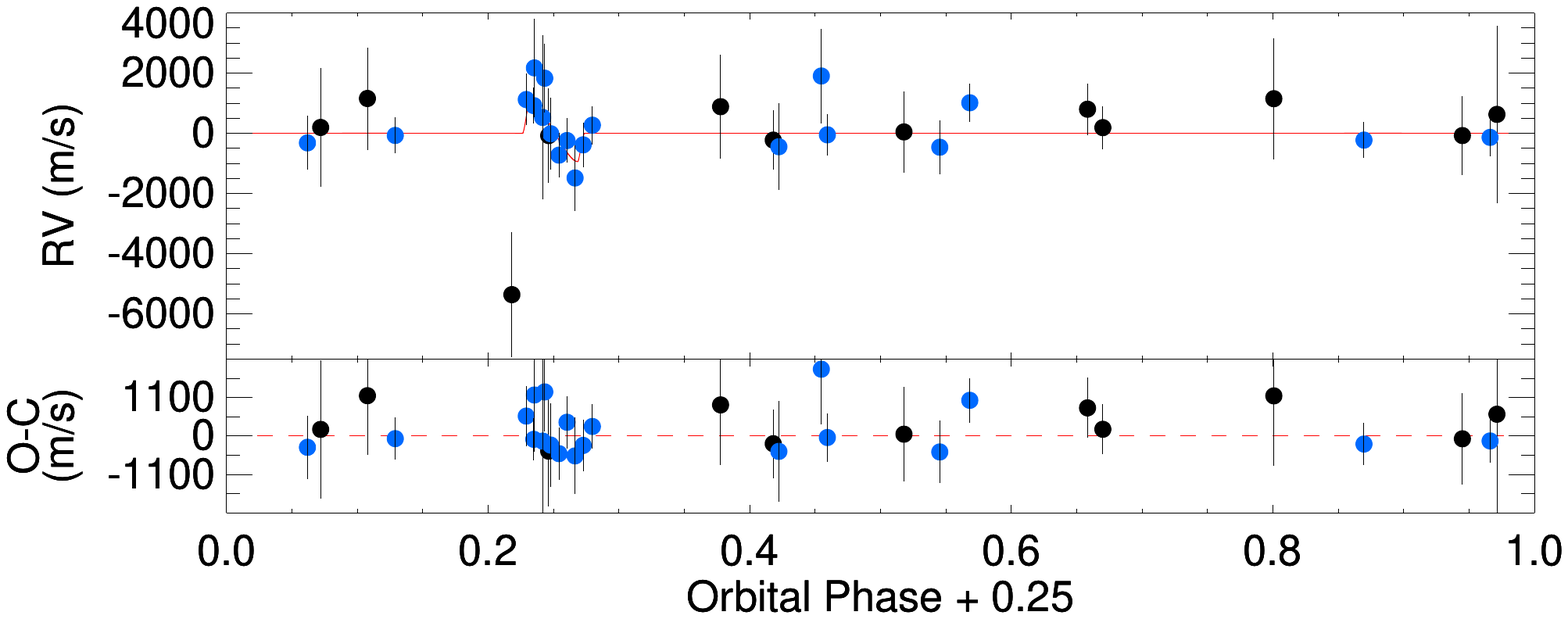}
   \vspace{.1in}
\caption{(Top) The TS23 (black) and TRES (blue) RV measurements of KELT-21 with the best-fit model shown in red. The systemic radial velocity has been subtracted from each dataset. The residuals to the fit are shown below. (Bottom) The RV measurements phase-folded to the global fit-determined ephemeris. The predicted radial velocity Rossiter-McLaughlin signal is shown at 0.25 phase. The residuals are shown below.}
\label{fig:RVs} 
\end{figure}

\subsubsection{TRES Spectra}
\label{sec:TRES}

We obtained 20 spectra of KELT-21 with the Tillinghast Reflector \'Echelle Spectrograph \citep[TRES; e.g.,][]{FureszThesis:2008}, on the 1.5\,m telescope at Fred Lawrence Whipple Observatory, Mt.\ Hopkins, Arizona, USA. TRES is a fibre fed \'echelle spectrograph with a spectral coverage of 3900--9100\,\AA{} over 51 \'echelle orders, with a spectral resolving power of $R=44,000$. The spectra are recorded by a $2048\times4608$ CCD.  We obtained 11 of the observations to constrain the mass of the planet, over the full range of out-of-transit phases, although we excluded two of these which happened to be obtained in-transit from the analysis. We also obtained nine spectra during the transit on 2017 May 29 UT to measure the Doppler tomographic signal of the transiting planet. These latter data are described further in \S\ref{sec:SpinOrbit}. Most of the observations had an exposure time of 1800 seconds, but a few had lengths of as short as 240 seconds or as long as 3000 seconds. The best TRES spectra have a per-pixel signal-to-noise ratio of $\sim65$ near 5200 \AA.

We reduced these data using version 2.55 of the TRES pipeline, a custom pipeline written in IDL by L.\ Buchhave. 
We analyzed the TRES spectra using two different methods, one in order to measure the absolute radial velocity of KELT-21, and another in order to measure the relative RVs of the star to constrain the planetary mass.

In order to determine the absolute velocity of KELT-21, we used only the six strongest out-of-transit TRES spectra. We cross-correlated these with a rotating model template spectrum with parameters similar to that of KELT-21 (see \S\S\ref{sec:SpecPars} and \ref{sec:GlobalFit}). The weighted average and standard deviation derived from these six spectra \citep[following][]{Buchave:2010,Quinn:2012} is $-12.4 \pm 1.0$ \kms\, on the TRES native system; converting to the IAU velocity system, this gives an absolute RV of $-13.0 \pm 1.0$ \kms\, for KELT-21.

The relative radial velocities were obtained by fitting the line broadening profiles of each spectrum. The line broadening profiles are derived via a least-squared deconvolution approach, as per \citet{Donati1997}, against a non-rotating synthetic template derived using the SPECTRUM\footnote{http://www1.appstate.edu/dept/physics/spectrum/spectrum.html} code with the ATLAS9 atmosphere models \citep{Castelli:2004}, over the wavelength range of 3900--6250\,\AA{}. The broadening kernel is fitted with a function accounting for rotation, macroturbulence, and instrumental broadening, and shifted in velocity to match the observation. The uncertainties were determined from the order-to-order velocity scatter, divided by the square root of the number of orders used.

We show these data in Fig.~\ref{fig:RVs}, and list the RV measurements in Table~\ref{tab:Spectra}.  

\subsubsection{TS23 Spectra}

We obtained 12 spectra of KELT-21 covering a wide range of orbital phases using the 2.7 m Harlan J. Smith Telescope at McDonald Observatory, Texas, USA, and its Robert G. Tull coud\'e spectrograph \citep{Tull:1995}. We conducted these observations between 2016 Oct 25 and 2017 Jun 19 UT. Observing conditions ranged from clear to thin or scattered clouds, with seeing typically between 1\farcs0 and 1\farcs5 but occasionally as poor as 2\farcs0. 

We used the spectrograph in its TS23 configuration, with a 1\farcs2$\times$8\farcs2 slit providing a resolving power of $R=60,000$ and coverage from 3570 \AA\,to 10200 \AA; the spectral coverage is complete below 5691~\AA. A $2048\times2048$ Tektronix CCD captures 58 spectral orders. Exposure times ranged from 120 to 1200 seconds, giving per-pixel signal-to-noise ratios as high as 120 (more typically $\sim60$) near 5500 \AA. We reduced the spectra using standard IRAF tasks, and measured relative radial velocities from the spectra using the same methodology as we used for the TRES data. The resulting measurements are shown in Fig.~\ref{fig:RVs} and tabulated in Table~\ref{tab:Spectra}.

\subsubsection{PEPSI Spectra}

We obtained high-resolution spectra of KELT-21 with $R=\lambda/\Delta\lambda=120,000$ with the Potsdam \'Echelle Polarimetric and Spectroscopic Instrument \citep[PEPSI;][]{Strassmeier:2015} spectrograph at the 2$\times$8.4\,m Large Binocular Telescope (LBT) on Mt.\ Graham, Arizona, USA. PEPSI is a fiber-fed white-pupil \'echelle spectrograph with two arms (blue and red optimized).  We employed two of the six cross dispersers (CD\,II and CD\,IV), which covered the wavelength ranges 4265--4800\,\AA\ and 5441--6278\,\AA\ simultaneously. The instrument is stabilized in a pressure- and temperature-controlled chamber and is fed by three pairs of octagonal fibers per LBT unit telescope \citep[for overall performance characterization we refer to][]{Strassmeier:2017}. For the present observations, we used the 200\,$\mu$m core fibers and the five-slice image slicer to achieve the spectral resolution of 120,000. The spectra are sampled with 4.2 pixels per resolution element. The fiber core projection on the sky is 1\farcs5. The spectrum is recorded by two 10.3k$\times$10.3k STA1600LN CCDs with 9 $\mu$m pixels. 

 The observations occurred on 2017 May 29 UT and lasted for 4 hours, concluding when we had to close for sunrise. We set the integration time to 1200~s. CCD read-out and overhead sums to 90~s and enabled a sequence of 13 back-to-back spectra. The target altitude was very low at the beginning of the observing sequence with an airmass of $\sim2.1$. The sky was clear and seeing at the start of the sequence was 1\farcs0 but deteriorated to 1\farcs3 at the end of the sequence. Peak signal-to-noise ratios were 140 and 100 per pixel in CD\,IV and CD\,II, respectively. 
 
 Data reduction was done with the software package SDS4PEPSI (``Spectroscopic Data Systems for PEPSI''), based on \cite{Ilyin:2000} and described in more detail in \cite{Strassmeier:2017}. It relies on adaptive selection of parameters by using statistical inference and robust estimators. The standard reduction steps include bias overscan detection and subtraction, scattered light extraction from the inter-order space and subtraction, definition of \'echelle orders, optimal extraction of spectral orders, wavelength calibration, and a self-consistent continuum fit to the full 2D image of extracted orders. Our Doppler tomographic analysis of these spectra is described in \S\ref{sec:SpinOrbit}.

\begin{table}
\centering
 \caption{Radial Velocities of KELT-21}
 \label{tab:Spectra}
 \begin{tabular}{lrrrr}
    \hline
    \hline
    \multicolumn{1}{l}{\bjdtdb} & \multicolumn{1}{c}{RV} 	& \multicolumn{1}{c}{$\sigma_{\rm RV}$} 	& \multicolumn{1}{c}{Phase} & \multicolumn{1}{c}{Instrument}\\
    & \multicolumn{1}{c}{(\kms)} &\multicolumn{1}{c}{(\kms)} & & \\
    \hline                                                          
2456902.88217 & -8.88 & 1.31 & 0.20 & TRES \\ 
2457642.70515$^{\dagger}$ & -9.87 & 0.50 & 0.98 & TRES \\ 
2457642.73542$^{\dagger}$ & -8.96 & 0.94 & 0.99 & TRES \\ 
2457902.80358$^{\dagger}$ & -9.67 & 0.71 & 0.98 & TRES \\ 
2457902.82614$^{\dagger}$ & -8.61 & 1.36 & 0.99 & TRES \\ 
2457902.84897$^{\dagger}$ & -10.26 & 2.28 & 0.99 & TRES \\ 
2457902.87114$^{\dagger}$ & -10.80 & 1.00 & 1.00 & TRES \\ 
2457902.89397$^{\dagger}$ & -11.51 & 0.62 & 0.00 & TRES \\ 
2457902.91618$^{\dagger}$ & -11.03 & 0.62 & 0.01 & TRES \\ 
2457902.93846$^{\dagger}$ & -12.27 & 0.91 & 0.02 & TRES \\ 
2457902.96073$^{\dagger}$ & -11.17 & 0.61 & 0.02 & TRES \\ 
2457903.94587 & -11.25 & 0.75 & 0.30 & TRES \\ 
2457905.81193 & -11.11 & 0.75 & 0.81 & TRES \\ 
2457910.86090 & -10.84 & 0.58 & 0.21 & TRES \\ 
2457913.82452 & -10.52 & 0.53 & 0.03 & TRES \\ 
2457914.86709 & -9.77 & 0.53 & 0.32 & TRES \\ 
2457915.95596 & -11.01 & 0.50 & 0.62 & TRES \\ 
2457916.89286 & -10.86 & 0.50 & 0.88 & TRES \\ 
2457917.95251 & -11.23 & 1.21 & 0.17 & TRES \\ 
2457919.91707 & -10.92 & 0.52 & 0.72 & TRES \\ 
2457686.57439 & -9.75 & 1.62 & 0.13 & TS23 \\ 
2457686.71975 & -10.86 & 0.93 & 0.17 & TS23 \\ 
2457687.58795 & -9.83 & 0.81 & 0.41 & TS23 \\ 
2457732.56483 & -9.48 & 1.60 & 0.86 & TS23 \\ 
2457734.59627 & -10.45 & 0.68 & 0.42 & TS23 \\ 
2457735.58946 & -10.71 & 1.23 & 0.69 & TS23 \\ 
2457736.57659 & -15.99 & 1.96 & 0.97 & TS23 \\ 
2457919.93596 & -10.01 & 2.78 & 0.72 & TS23 \\ 
2457920.92942 & -10.71 & 1.48 & 1.00 & TS23 \\ 
2457921.91076 & -10.59 & 1.27 & 0.27 & TS23 \\ 
2457922.93250 & -9.49 & 1.89 & 0.55 & TS23 \\ 
2457923.91227 & -10.44 & 1.86 & 0.82 & TS23 \\ 
\hline
 \end{tabular}
 \begin{flushleft}
  \footnotesize{ \textbf{\textsc{NOTES:}}  See Table~\ref{tbl:KELT-21b} for the RV zero point values. We have assumed a minimum uncertainty of 0.5 \kms\ on the radial velocities measurements. $^{\dagger}$: denotes that the observation takes place during transit and was excluded from our RV analysis (although the spectra from BJD 2457902 were used in our Doppler tomographic analysis: \S\ref{sec:SpinOrbit}).}
  
\end{flushleft}
\end{table} 

\subsection{High Contrast AO Imaging}
\label{sec:AO}

\begin{figure}
\centering 
\includegraphics[width=1.05\columnwidth, trim = 0 0.75in 3.5in 1.0in]{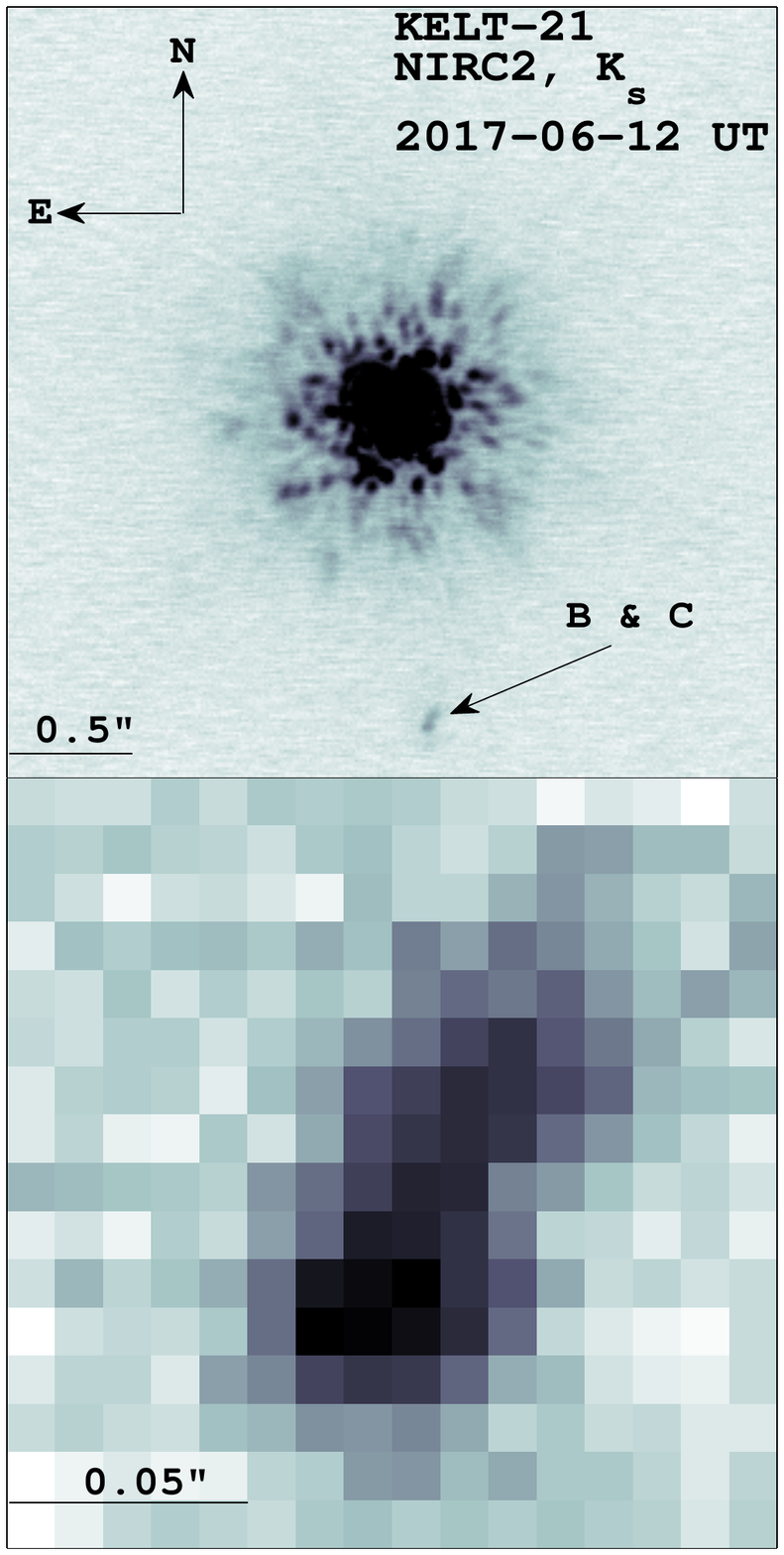}\\
\vspace{-36pt}
\includegraphics[width=1.05\columnwidth]{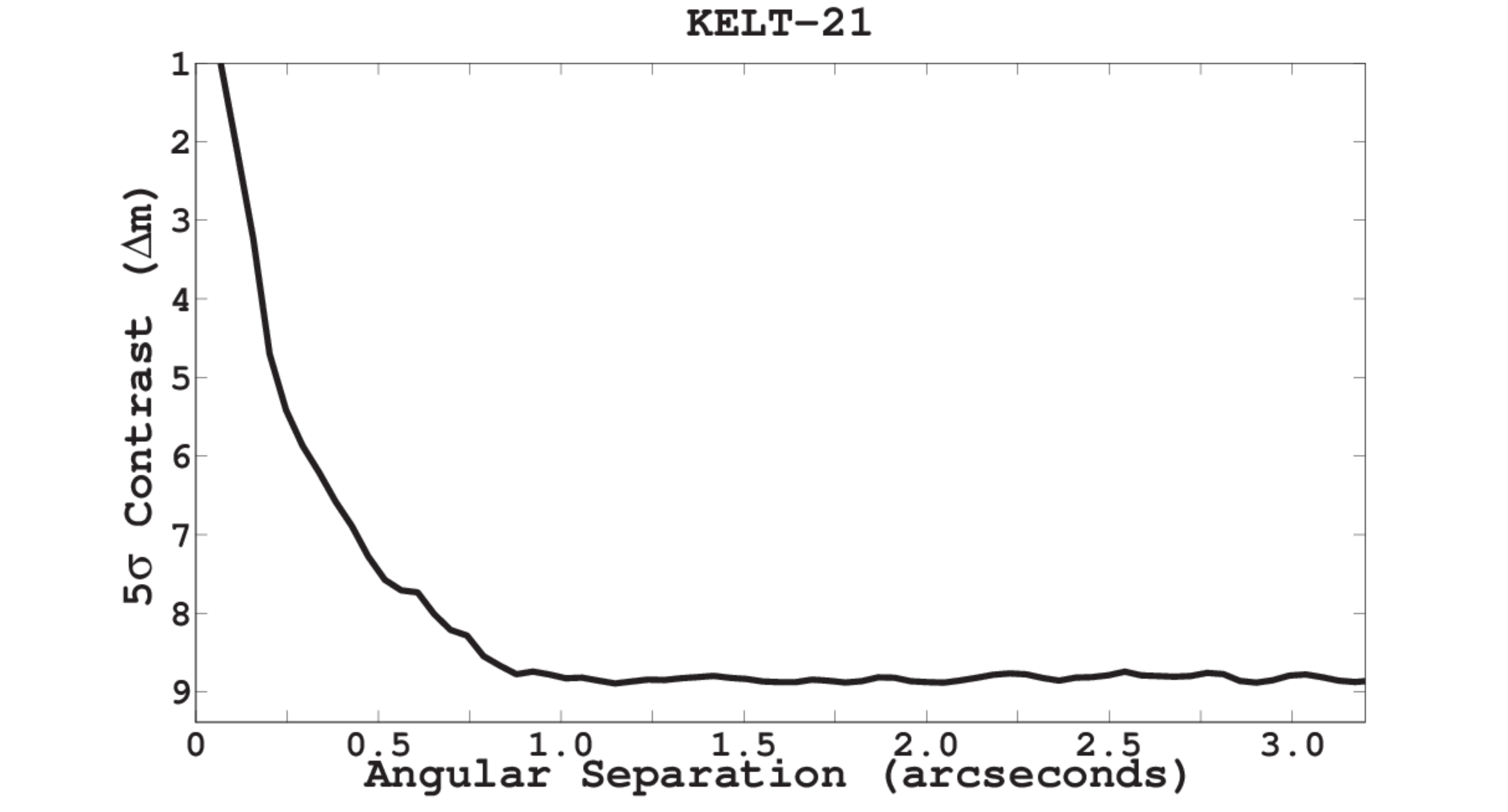}
\caption{\footnotesize Top: Keck NIRC2 AO image of the KELT-21 system, showing the primary star and the two faint companions B and C. Middle: zoom-in on the companions B and C. Bottom: contrast curve for these observations.}
\label{fig:AO}
\end{figure}

We obtained adaptive optics (AO) images of KELT-21 on 2017 June 12 UT using the NIRC2 instrument on Keck II, Maunakea, Hawaii, USA. The observing conditions were excellent. KELT-21 was observed at an airmass of $\sec{z}=1.07$ with seeing estimated to be $\sim$0\farcs3. The narrow camera mode was used to provide a plate scale of $9.942 \pm 0.05$ mas pix$^{-1}$. A three-point dither pattern was implemented to avoid the noisy quadrant of the NIRC2 detector. A total of thirty frames were recorded using position angle mode in the $K_s$ band; the sequence resulted in a total integration time of thirty seconds. No off-axis sources were identified in raw frames, so we did not obtain images in complementary filters. Later inspection of processed frames, however, revealed two faint companions to the south of KELT-21 (top panel of Fig.~\ref{fig:AO}).     

Standard AO imaging reduction methods were used to flat-field the images, correct for bad pixels, and subtract the sky background \citep{Crepp:2012}. Upon noticing the companions from the NIRC2 automated pipeline, we performed several experiments to confirm their nature, including studying how the signal-to-noise ratio improved with increasing number of processed frames from different detector quadrants. We also carefully assessed image registration internal to the pipeline to ensure that virtual copies of the primary star or other effects were not creating multiple off-axis signals. Finally, a contrast curve was generated to serve as a self-consistency check for the detected companions' flux levels and limit the existence of other sources near KELT-21 (bottom panel of Fig~\ref{fig:AO}). In order to construct the contrast curve, we divide the image into a grid with a cell size set to the FWHM of the PSF, and calculate the RMS over each 3$\times$3 set of cells. The contrast curve denotes the 5$\sigma$ median-combined RMS value azimuthally averaged over the $3\times3$ cells centered at a given radial separation from KELT-21. The companions are more than a magnitude above the contrast curve, indicating that their detection is secure, and we do not detect any other sources of comparable or greater brightness near KELT-21. 

Although well-separated from the primary star by just over an arcsecond, the two sources themselves have comparable flux and are only marginally spatially resolved. Utilizing brute-force modeling of the companions as a combined source using aperture photometry, we find a combined magnitude difference of $\Delta K_s = 6.39 \pm 0.06$ compared to the primary star. 
We then employed Markov Chain Monte Carlo (MCMC) methods to compute the relative brightness and astrometric positions of the closely separated binary pair. To disentangle the contribution of each source, we use the technique described in \citet{Bechter:2014} to self-consistently fit the AO data. Specifically, we model the core and halo of each source using a modified Moffat function that includes nuisance parameters such as the residual sky background pedestal. After numerically identifying the multi-dimensional parameter space global minimum, we ran 1 million iterations to generate posterior distributions that quantify the relative position and brightness of the putative companions and their uncertainties. Results are shown in Table~\ref{tab:astr_phot}. We will consider the candidate companions in more detail in \S\ref{sec:Triple}.

\begin{table}
\centering
 \caption{Properties of the Likely Companions KELT-21 B and C}
 \label{tab:astr_phot}
 \begin{tabular}{l l c}
    \hline
    \hline
    \multicolumn{1}{l}{Parameter} & \multicolumn{1}{l}{Description (Units)} 	& \multicolumn{1}{c}{Value}  \\
    \hline
\multicolumn{3}{l}{KELT-21 B} \\
$\rho_{AB}$ &  Separation (mas) & $1261 \pm 12$ \\
$PA_{AB}$ & Position Angle ($^{\circ}$, east of north) & $185.3 \pm 0.1$ \\
$\Delta K_s$ & Contrast (mag) &  $7.00 \pm 0.06$ \\
$a_{\perp,AB}$ & Projected Separation (AU) & $523 \pm 62$ \\
$M_B$ & Estimated Mass (\msun) & $0.13_{-0.01}^{+0.02}$ \\
\hline
\multicolumn{3}{l}{KELT-21 C} \\
$\rho_{AC}$ &  Separation (mas) & $1214 \pm 14$ \\
$PA_{AC}$ & Position Angle ($^{\circ}$, east of north) & $186.6 \pm 0.1$ \\
$\Delta K_s$ & Contrast (mag) &  $7.30 \pm 0.06$ \\
$a_{\perp,AC}$ & Projected Separation (AU) & $504 \pm 60$ \\
$M_C$ & Estimated Mass (\msun) & $0.11 \pm 0.01$ \\
\hline 
\multicolumn{3}{l}{KELT-21 BC} \\
$\rho_{BC}$ &  Separation (mas) & $55 \pm 16$ \\
$PA_{BC}$ & Position Angle ($^{\circ}$, east of north) & $324 \pm 10$ \\
$\Delta K_s$ & Contrast (mag) &  $0.30 \pm 0.08$ \\
$a_{\perp,BC}$ & Projected Separation (AU) & $22.9 \pm 7.1$ \\
    \hline
    \hline
 \end{tabular}
  \begin{flushleft}
  \footnotesize{Note: all parameters are quoted at the epoch of the observations, 2017 Jun 12 UT. Subscripts AB, AC, and BC refer to mutual parameters between stars A (the planet host star KELT-21) and B, A and C, and B and C, respectively.
  The physical parameters of KELT-21 B and C are calculated assuming that they are physically associated with KELT-21; see \S\ref{sec:Triple}. The quoted uncertainties on these parameters are calculated given the formal uncertainties on the parameters of KELT-21 and the photometric measurements but neglect sources of systematics such as uncertainties in stellar isochrones.}
\end{flushleft}
\end{table}

\section{Host Star Characterization}
\label{sec:Star}

\subsection{SED Analysis}
\label{sec:SED}

We followed the approach of previous KELT discoveries and fitted the broadband spectral energy distribution (SED) of KELT-21 using a \citet{Kurucz:1992} atmosphere model. We adopted the broadband fluxes from the available all-sky photometric catalogs, in particular $B_T$ and $V_T$ from {\it Tycho-2}, $JHK_S$ from {\it 2MASS}, and WISE1--3 from {\it AllWISE}. We also adopted the 
{\it Gaia\/} DR1 parallax, $\pi = 2.41 \pm 0.28$~mas, corrected for the systematic offset determined by \citet{Stassun:2016}. These data are listed in Table~\ref{tab:LitProps}.
The free parameters of the fit were \teff\ and the extinction, $A_V$, limited by the maximum line-of-sight $A_V$ from the \citet{Schlegel:1998} dust maps. Since \loggstar\ and \feh\ are of secondary importance to the SED, we assumed a main-sequence \loggstar\ = 4 and solar metallicity. We neglected the presence of the candidate companions discussed in \S\ref{sec:AO} as they are much fainter than KELT-21 (combined $\Delta K_S=6.39 \pm 0.06$) and so should have a negligible effect on the SED.

The best-fit SED, has $\chi_\nu^2 = 4.2$ for 5 degrees of freedom. 
The best-fit parameters are 
\teff\ = $8000_{-250}^{+1000}$~K and $A_V = 0.00_{-0.00}^{+0.34}$.
Integrating the SED gives an extinction-corrected bolometric flux at Earth of $F_{\rm bol} = 1.61_{-0.13}^{+0.78} \times 10^9$ ergs s$^{-1}$ cm$^{-2}$.

\subsection{Spectroscopic Analysis}
\label{sec:SpecPars}

We determined the properties of KELT-21 using The Payne, a newly developed approach for determining stellar parameters from simultaneously fitting the observed spectrum and spectral energy distribution self-consistently with {\it ab initio} models. The basic framework of The Payne algorithm is given in \cite{Ting:2016} and \cite{Rix:2016}, and full details of the code will be given in Cargile et al.\ (in prep). 

Using The Payne, we model five of the individual TRES spectra of KELT-21. In the inference, we fit a wavelength region of $\sim$200 \AA\ around the Mg \textsc{i} (Mg b) triplet at $\sim$5200 \AA, and all available photometry from the Tycho-2, 2MASS, and AllWISE catalogs (Table~\ref{tab:LitProps}); we did not use the AllWISE $W4$ band as there is only a limit available in $W4$. For each TRES spectrum, we infer the most probable \teff, \loggstar, \feh, [$\alpha$/Fe], radial velocity, intrinsic stellar broadening, instrumental broadening profile, stellar radius, distance, and extinction in the $V$ band (A$_{V}$). We apply priors on the known instrumental profile for the TRES instrument ($R=44,000$), the Gaia DR1 parallax distance ($\Pi =2.41 \pm 0.28$ mas), and the surface gravity inferred from the planetary transit model (\loggstar$=$4.16$\pm$0.1; see \S\ref{sec:GlobalFit}). 

\begin{figure}
\includegraphics[width=1.1\columnwidth]{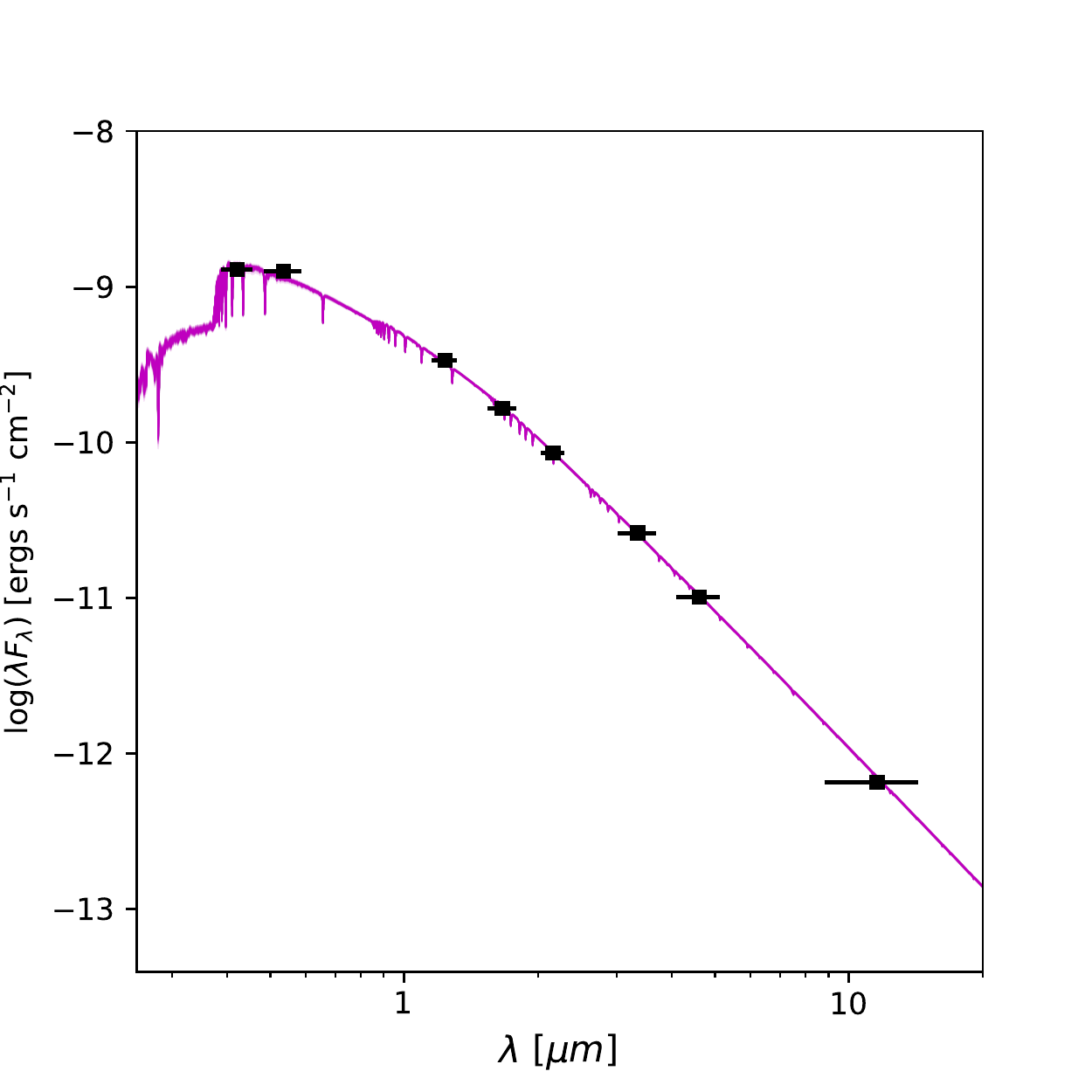}
\caption{\footnotesize Spectral energy distribution of KELT-21. The purple line shows the best-fit model from The Payne and 50 random draws from the posteriors (only visible as a finite width to the line), and the black points show the literature photometry used. The error bars on each point represent the width of each photometric band.}
\label{fig:SED}
\end{figure}

In order to determine the overall best-fit stellar parameters for KELT-21, we take the median and standard deviation of the results from the modeling of the five individual spectra. We note that the standard deviation of the five TRES spectra is very similar to the measurement errors on the most probable fit for the individual spectra, suggesting that it is a good representation of the formal measurement uncertainties from The Payne. We show the best-fit SED in Fig.~\ref{fig:SED}.

\begin{figure}
\includegraphics[width=1.1\columnwidth]{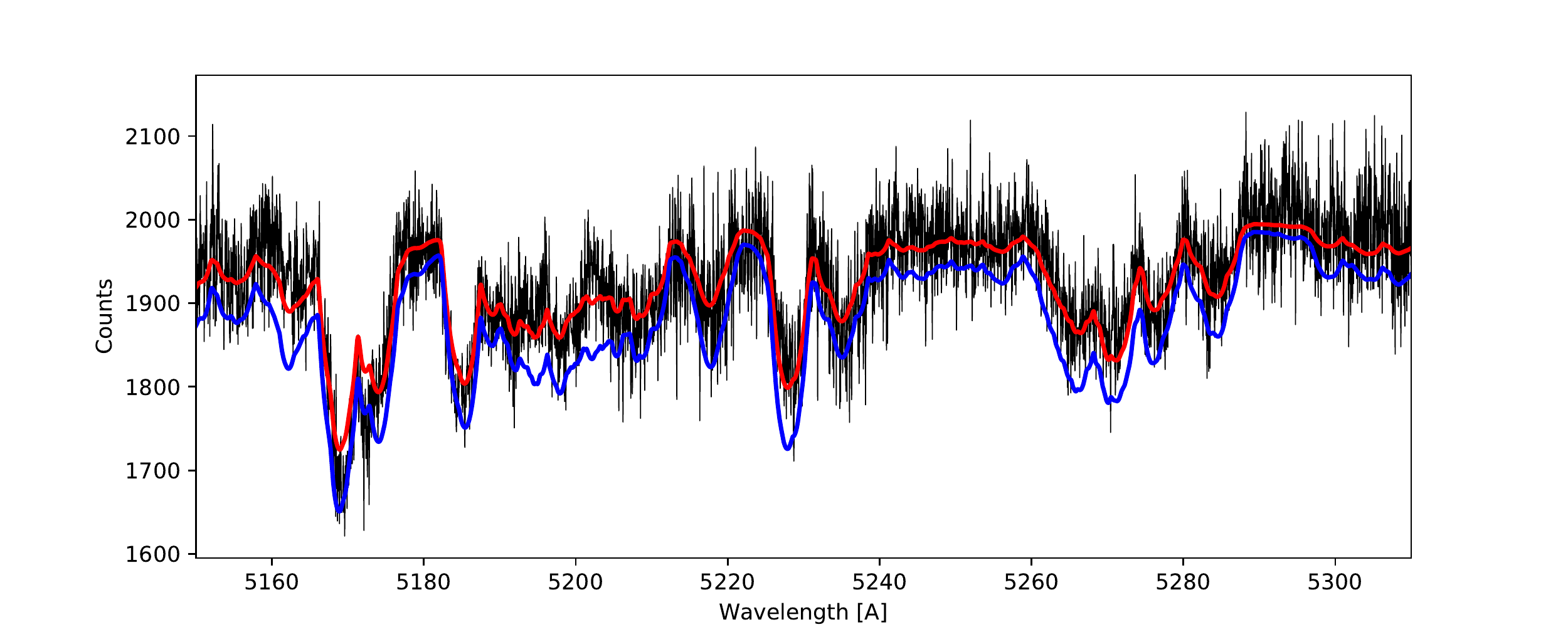}
\caption{\footnotesize A section of one TRES spectrum of KELT-21, showing the data in black, the best-fit model from The Payne (with \feh=$-0.410 \pm 0.032$, [$\alpha$/Fe]=$0.145 \pm 0.053$) in red, and a model with the same stellar parameters except \feh=0.0, [$\alpha$/Fe]=0.0 in blue. It is visually apparent that the low-metallicity, $\alpha$-enhanced model is a better fit to the data--the solar-metallicity model generally overpredicts the line depths--and so we conclude that the low metallicity of KELT-21 is robust despite the rapid stellar rotation.}
\label{fig:FeHcomp}
\end{figure}

Our analysis with The Payne produced stellar parameters of \teff=$7587 \pm 82$ K, \feh=$-0.410 \pm 0.032$, [$\alpha$/Fe]=$0.145 \pm 0.053$, and \vsinistar$=144.3 \pm 1.2$ \kms. The uncertainties on our metallicity and $\alpha$-enhancement measurements are small, which is somewhat surprising given the difficulty of spectral analysis for hot, rapidly rotating stars like KELT-21. Nonetheless, we argue the sub-solar metallicity is robust. In Fig.~\ref{fig:FeHcomp} we show part of one of our TRES spectra, along with the best-fit model and a model with solar metallicity. The solar-metallicity model consistently overpredicts the depths of the spectral lines. Additionally, solutions with \feh=0, [$\alpha$/Fe]=0 have posterior probabilities of $\sim10^{-8}-10^{-10}$ (as do solutions with either \feh=0 or [$\alpha$/Fe]=0 individually) in most of our fits with The Payne. We thus conclude that KELT-21 is indeed metal-poor, which is unusual for relatively young ($<2$ Gyr), hot stars; we explore the implications of this measurement in more detail in \S\ref{sec:metals}. Additionally, the \teff\ value found here is 1.6$\sigma$ discrepant from that found in our SED analysis in \S\ref{sec:SED}. This is likely due to the fact that our SED-only analysis assumed \feh$=0$, which is not the case; our analysis with The Payne provides an equally good fit to the literature photometry, and so we proceed using these stellar parameters. The Payne value of \vsinistar is mostly consistent ($1.3\sigma$ difference) with that measured independently from the PEPSI spectra (\S\ref{sec:GlobalFit}).

A \teff\ of 7587 K corresponds to a spectral type of A8, per the \teff-spectral type calibration of \cite{Pecaut:2013}, while the surface gravity value of \loggstar=$4.173_{-0.015}^{+0.016}$ from the global fit (\S\ref{sec:GlobalFit}) indicates that KELT-21 is on the main sequence. We therefore find a spectral type of A8V for KELT-21; this is one of only a handful of A stars known to host a transiting planet.

\subsection{Evolutionary Analysis}
\label{sec:Evol}

\begin{figure}
\includegraphics[width=1.0\columnwidth, trim = 1.0in 1.0in 1.0in 1.0in]{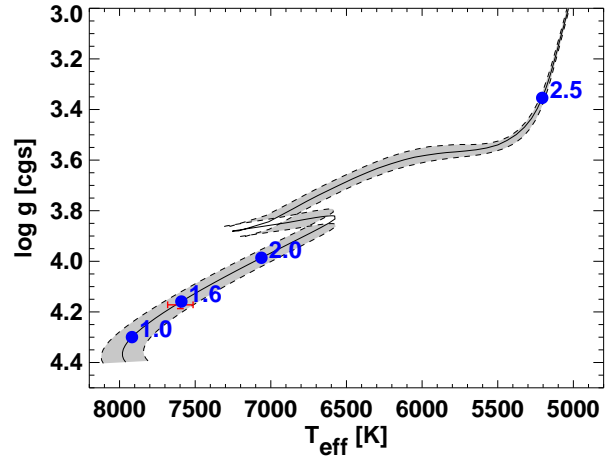}
\caption{The location of KELT-21 in the Kiel diagram. The best-fit \teff\ and \loggstar\ from the global model fit is shown as the red point, while the gray swath shows a Yonsei-Yale evolutionary track for a star with the best-fit values of \mstar\ and \feh; the locations on the best-fit model corresponding to several values of the stellar age are shown as the blue points, with ages quoted in Gyr.} 
\label{fig:hrd}
\end{figure}

In Fig.~\ref{fig:hrd} we show KELT-21 in the \teff-\loggstar plane, i.e., the Kiel diagram, along with a Yonsei-Yale \citep[YY;][]{Demarque:2004} evolutionary track for the mass and \feh\ of KELT-21. The evolutionary track implies an age of $\sim1.6 \pm 0.1$ Gyr for KELT-21. KELT-21 will start evolving off the main sequence within the next few hundred million years, and within $\sim1$ Gyr it will begin ascending the red giant branch.

\subsection{UVW Space Motion and Galactic Location}
\label{sec:UVW}

In order to put the KELT-21 system into a broader Galactic context, we calculated its three-dimensional Galactic space motion ($U,V,W$). We obtained the proper motion and parallax of KELT-21 from Gaia DR1 \citep{Brown:2016}, and corrected the parallax for systematic biases per the formulation of \cite{Stassun:2016}. We also used the absolute radial velocity of the system on the IAU scale as determined from our TRES spectra (\S\ref{sec:TRES}). All of these parameters are listed in Table~\ref{tab:LitProps}. 

We calculated the Galactic space motion using the IDL routine \texttt{GAL\_UVW}\footnote{https://idlastro.gsfc.nasa.gov/ftp/pro/astro/gal\_uvw.pro}, which is based upon the methodology of \cite{Johnson:1987}, and we used the value of the Solar velocity with respect to the local standard of rest found by \cite{Coskunoglu:2011}. We find that KELT-21 has a space motion of ($U$,$V$,$W$) = ($9.2 \pm 1.9$, $-0.5 \pm 1.1$, $8.9 \pm 1.9$) km s$^{-1}$; we use the coordinate system such that the Galactic center is in the direction of positive $U$. Using the criteria of \cite{Bensby:2003}, this corresponds to a 99.4\% probability that KELT-21 is a member of the Milky Way's thin disk. This is expected given the relatively high mass, and therefore relatively low age, of KELT-21. We note that the \cite{Bensby:2003} criteria are not strictly applicable to KELT-21, as they were derived for the solar neighborhood and KELT-21 is located at a distance of 0.4 kpc. KELT-21 is located close to the solar circle, however (as $l=71.4814^{\circ}$, $b=-1.9865^{\circ}$), suggesting that the \cite{Bensby:2003} criteria are likely not unreasonable. Indeed, our integration of the orbit of KELT-21 (\S\ref{sec:metals}) confirms its thin-disk kinematics.

Given that KELT-21 is located very close to the Galactic plane, significant extinction and reddening might be expected. The Pan-STARRS 1 dust map\footnote{http://argonaut.skymaps.info/} \citep{Green:2015} predicts a reddening of $E(B-V)=0.09^{+0.02}_{-0.03}$ at the distance and position of KELT-21, although this is close enough that there are not enough stars for the dust map to be fully reliable. Using a standard value of $R_V=3.1$, this corresponds to an expected extinction of $A_V=0.28_{-0.09}^{+0.06}$, which is consistent to within $1\sigma$ with the value of $A_V=0.00_{-0.00}^{+0.34}$ found in our SED analysis (\S\ref{sec:SED}). The PEPSI spectra also show significant interstellar absorption with a complex velocity structure in the Na \textsc{i} D lines, as is expected for a star in this direction and distance.

\section{Planet Characterization}
\label{sec:Planet}

\subsection{Doppler tomographic characterization}
\label{sec:SpinOrbit}

We analyzed the PEPSI and in-transit TRES data using Doppler tomographic methodology. When a planet transits a rotating star, the obscured regions of the stellar disk do not contribute to the formation of the rotationally-broadened stellar absorption line profile. The subtracted light results in a perturbation to the line profile at velocities corresponding to the radial velocities of the obscured surface elements. This is known as the Rossiter-McLaughlin effect \citep{Rossiter:1924,McLaughlin:1924}. For slowly-rotating stars, this is typically interpreted as an anomalous radial velocity shift during the transit due to the changing line centroids \citep[e.g.,][]{Triaud:2010}. For sufficiently rapid rotation and/or sufficiently high spectral resolution, however, we can spectroscopically resolve the rotationally-broadened line profile and the line profile perturbation. This is Doppler tomography \citep[e.g.,][]{CollierCameron2010,Johnson:2014}. Detection of the line profile perturbation confirms that the planet candidate does indeed transit the target rapidly rotating star. Furthermore, the motion of the line profile perturbation across the line profile during the transit is diagnostic of the spin-orbit misalignment $\lambda$, which is the angle between the stellar spin and planetary orbital angular momentum vectors projected onto the plane of the sky.

Our procedure to extract the time series line profiles from the PEPSI data is essentially the same as used by \cite{Johnson:2014,Johnson:2015,Johnson:2017}. In short, we use least squares deconvolution \citep{Donati1997} to extract the average line profile from each spectrum. We use a line mask with initial guesses for the line depths taken from a Vienna Atomic Line Database \citep[VALD;][]{Ryabchikova:2015} stellar model with the stellar parameters of KELT-21; best-fit line depths are then found using the stacked PEPSI spectra before the final line profiles are extracted. 

\begin{figure*}
\centering 
\includegraphics[scale=0.52, trim = 1.0in 4.95in 0.5in 4.25in]{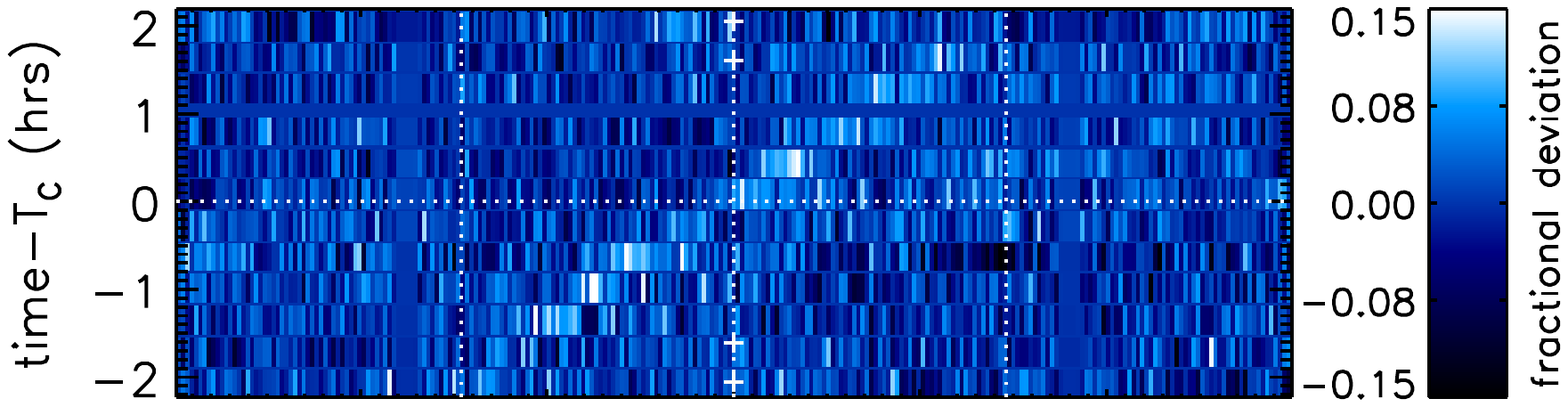}
\includegraphics[scale=0.52, trim = 1.0in 4.95in 1.0in 4.25in]{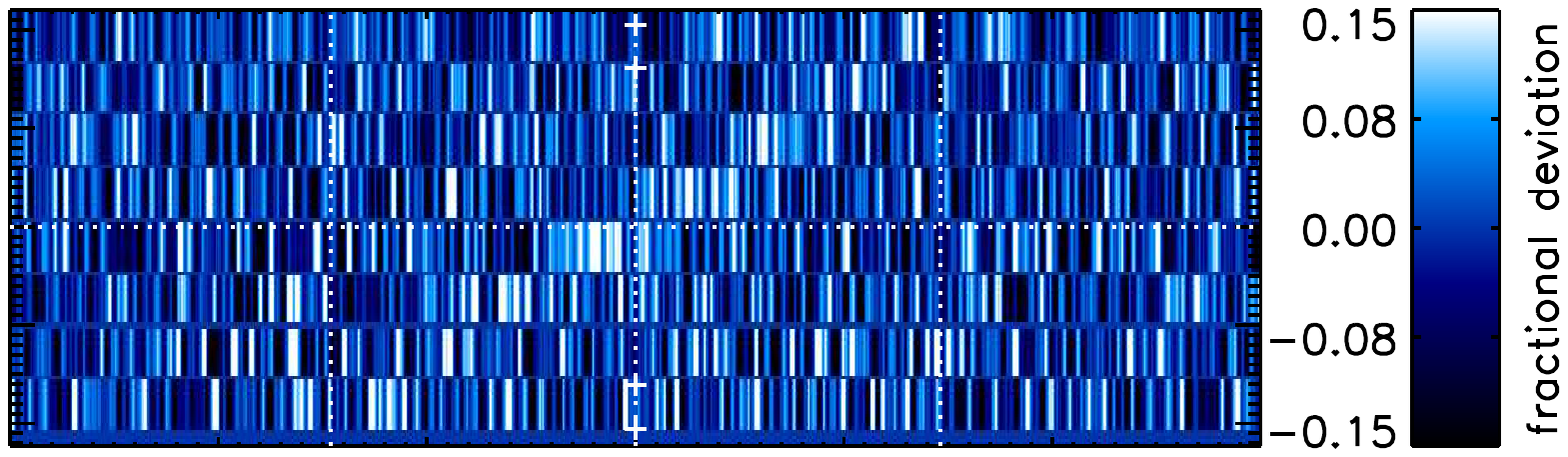} \\
\includegraphics[scale=0.52, trim = 1.0in 4.25in 0.5in 4.25in]{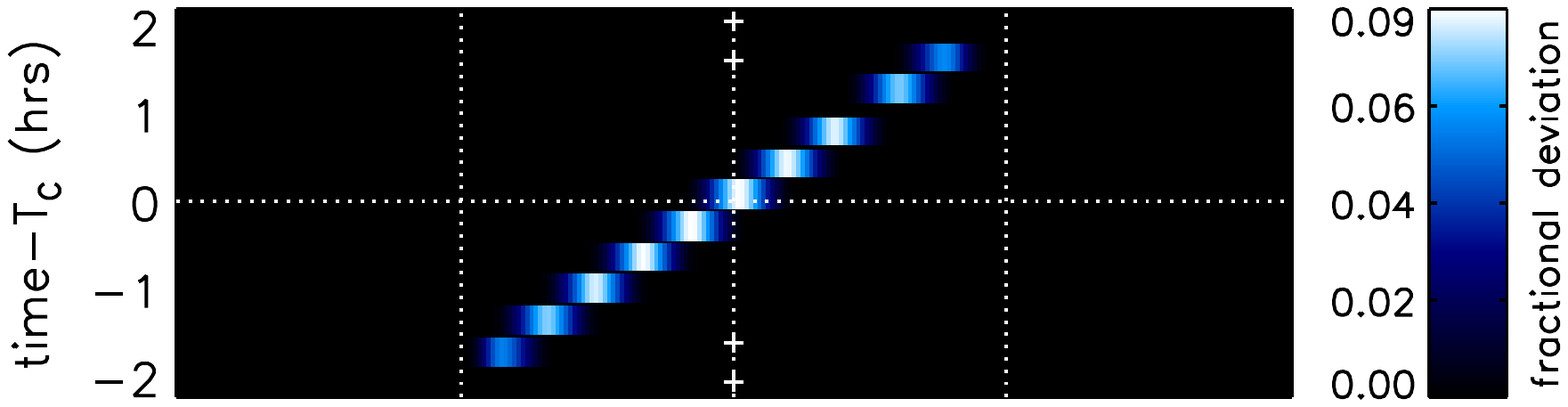}
\includegraphics[scale=0.52, trim = 1.0in 4.25in 1.0in 4.25in]{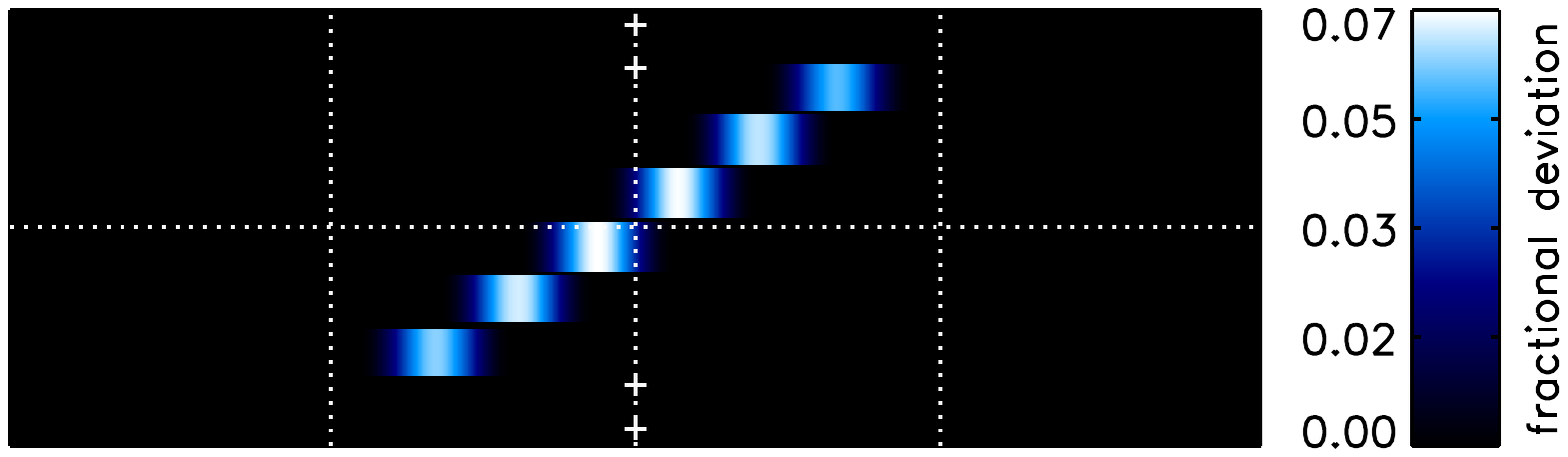} \\
\includegraphics[scale=0.52, trim = 1.0in 5.25in 0.5in 4.25in]{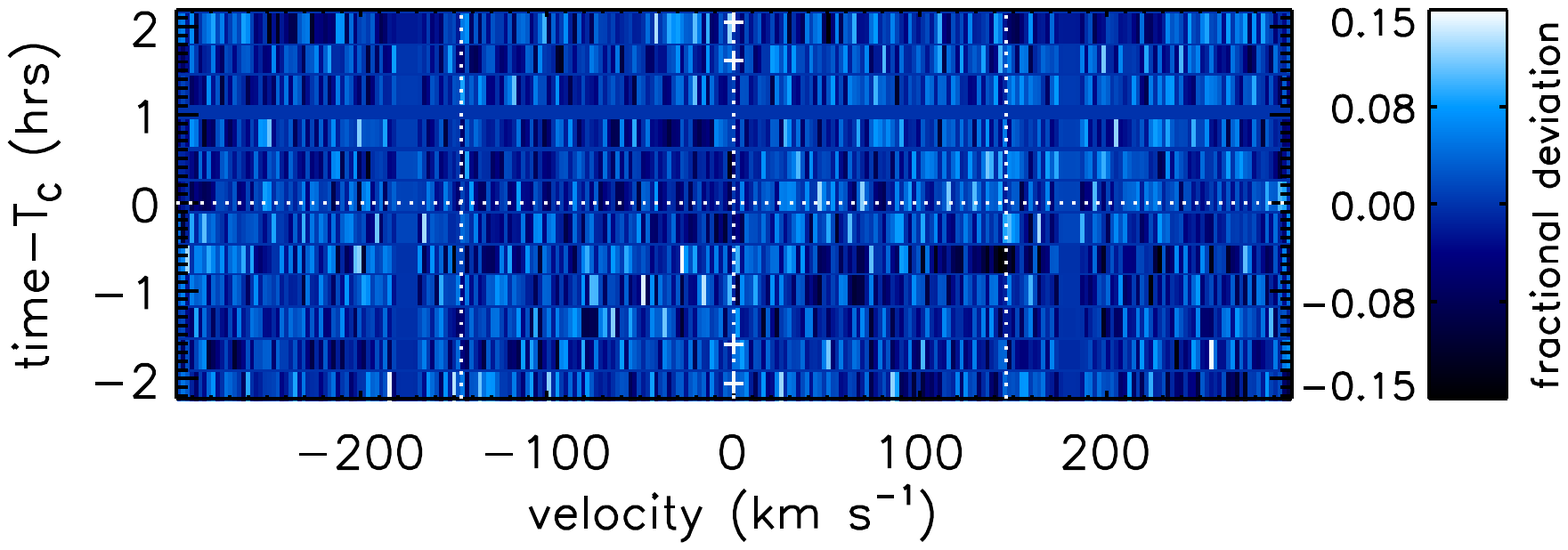}
\includegraphics[scale=0.52, trim = 1.0in 5.25in 1.0in 4.25in]{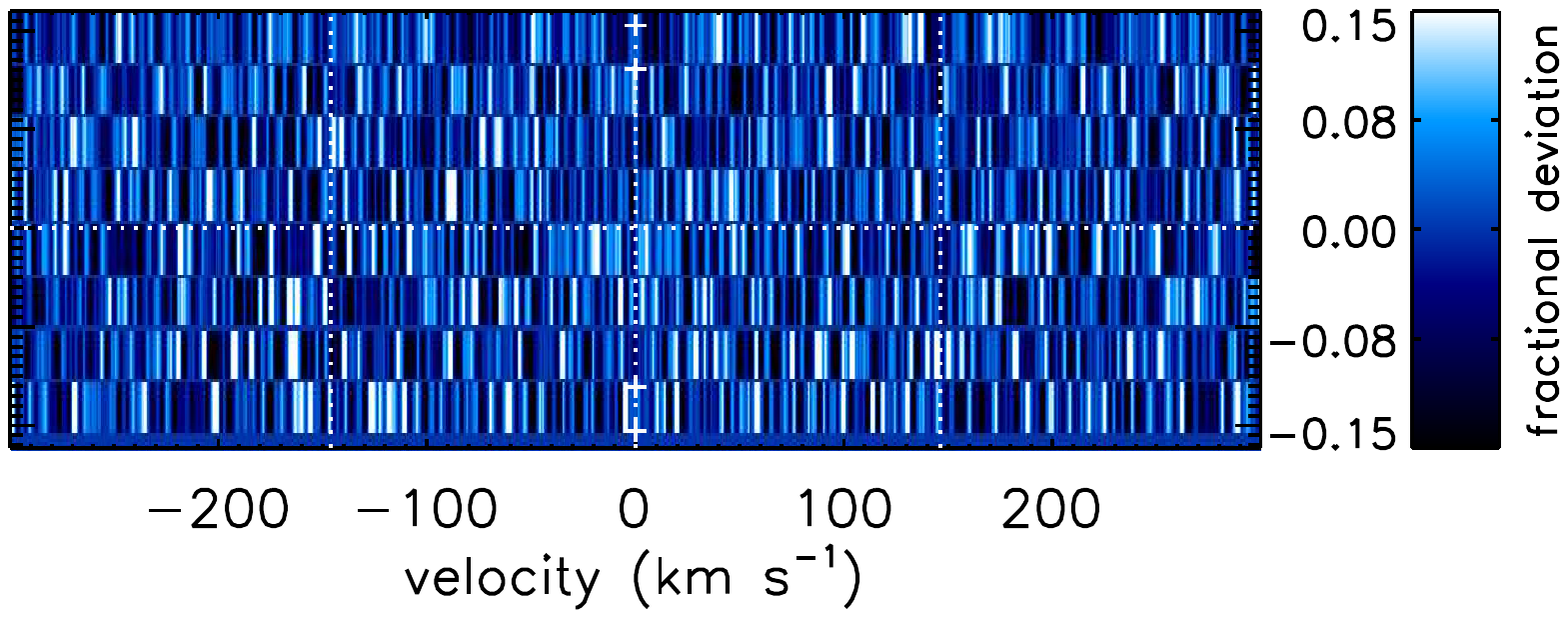} \\
\caption{\footnotesize The Doppler tomographic data for KELT-21. The PEPSI observations are shown in the left column, and the TRES observations in the right column. The top row shows the data, the middle row the best-fit model of the line profile perturbation due to the transit, and the bottom row the residuals after subtraction of the best-fit model. In all panels time increases from bottom to top, and each colorscale row shows the deviation of the line profile at that time from the average line profile. The transit is the bright streak moving from lower left to upper right. Vertical lines mark the center of the line profile at $v=0$ and the edges at $v=\pm\vsinistar$, a horizontal line shows the time of mid-transit, and the four small crosses depict the times of first through fourth contacts. The data have been binned down by a factor of 4 in the velocity axis as compared to the raw data for display purposes to better show the transit; the full-resolution data were used in the global fits. All panels showing data use the same color-scale range for better comparison; all pixels outside this range have been set to these extreme values. }
\label{fig:DT}
\end{figure*}

One modification from earlier methodology is necessary because of the narrower bandwidth of PEPSI with respect to the spectrographs used by \cite{Johnson:2014,Johnson:2015,Johnson:2017}. A significant fraction of the PEPSI blue arm spectrum is occupied by the H$\gamma$ line and its wings. Rather than entirely excluding this region of the spectrum, as was done in the vicinity of strong lines by \cite{Johnson:2014,Johnson:2015,Johnson:2017}, we instead subtract off a model of the H$\gamma$ line profile from \cite{Kurucz:1979}\footnote{As tabulated at http://kurucz.harvard.edu/grids/gridm01/bm01k2.datcd}. The PEPSI red arm spectrum has very few lines strong enough to use for the Doppler tomographic analysis; we only utilize a few lines blueward of 5680 \AA, while the Na \textsc{i} D lines are not usable as they suffer from strong interstellar contamination. We show the extracted time series line profile residuals, displaying the planetary transit, in the top left panel of Fig.~\ref{fig:DT}.

For the TRES Doppler tomographic observations, we calculate a line broadening kernel from each transit spectrum as per the least-squares deconvolution process described in Section~\ref{sec:TRES}. To detect the tomographic shadow of the transiting planet, we subtract the out-of-transit broadening kernel from the in-transit kernels \citep{CollierCameron2010,Zhou:2016}. The residuals, showing the transit signal, are shown in the upper right panel of Fig~\ref{fig:DT}.

By inspection it is apparent that the planetary orbit is prograde and well-aligned, as the line profile perturbation moves from the blueshifted to the redshifted limb during the course of the transit, and spans nearly the full velocity range of $\pm\vsinistar$. Our global fit (\S\ref{sec:GlobalFit}) confirms this qualitative assessment: we obtain $\lambda=-5.6_{-1.9}^{+1.7 \circ}$. The implications of this measurement are discussed in \S\ref{sec:lambda} and \S\ref{sec:compimplications}.

\subsection{EXOFAST Global Fit}
\label{sec:GlobalFit}
To measure the system parameters for KELT-21, we simultaneously fit all photometric and spectroscopic (including Doppler tomographic) data using a heavily modified version of EXOFAST \citep{Eastman:2013}. To determine the radius and mass of KELT-21, we use the YY stellar evolutionary tracks \citep{Demarque:2004} or the Torres empirical relations \citep{Torres:2010}. We conduct two separate global fits (YY and Torres) with the eccentricity of the planet's orbit set to zero. We assume a circular orbit as our radial velocity measurements are unable to measure the planetary mass, let alone provide meaningful constraints on the orbital eccentricity, and hot Jupiters also typically have circular or nearly circular orbits due to strong tidal damping. For a detailed description of the global modeling process, see \citet{Siverd:2012}. We use the SED + spectroscopic fit determined \teff\ of $7587 \pm 82$ K and \feh\ of $-0.410 \pm 0.032$ (\S\ref{sec:SpecPars}) as a prior. From an analysis of the KELT-North light curve, we add a prior on the period and transit center time. We also included priors of \vsinistar$=146.0 \pm 0.5$ \kms\ and a width of a Gaussian non-rotating line profile of $5.2 \pm 0.8$ \kms\, derived from a preliminary fit to the PEPSI line profiles (\S\ref{sec:SpinOrbit}) using the line profile model described in \cite{Johnson:2014}. 

After the global fit (both YY and Torres) we measure an independent ephemeris from a linear fit of the determined transit center times for each follow up light curve. We then reran both fits using this new $T_c$ and Period and their uncertainties as priors to obtain the final results. We do not include the KELT-North light curve in any of the global fits due to difficult-to-quantify blending. We also place a prior on \rstar\ of 1.53$\pm$0.43 \rsun\ using the {\it Gaia} parallax (See Table \ref{tab:LitProps}) and the measured bolometric flux from our SED analysis (\S\ref{sec:SED}). We adopt the YY fit for the discussion of the KELT-21 system. See Tables \ref{tbl:KELT-21b} and \ref{tbl:KELT-21b_part2} for the results of both global fits. 

We neglected any contribution of the candidate companions described in \S\ref{sec:AO} to the photometric data or the spectra. Based upon the $\Delta K_S$ contrast values from the AO data and assuming physical association of the companions with KELT-21 (see \S\ref{sec:Triple}), we estimate flux ratios between the exoplanet host star and the combined light of the companions of $1.5\times10^{-4}$ in the $V$ band and $3.3\times10^{-4}$ in the $I_C$ band. Not accounting for this flux would cause a systematic underestimate in the transit depth much less than the uncertainty due to photometric noise, and so we can safely neglect this contaminating flux. We also note that this implies that the companions are too faint to cause the transit signal if one of them were to be an eclipsing binary (a possibility which is also excluded by our detection of the Doppler tomographic transit signal).

\begin{table*}
 \scriptsize
\centering
\setlength\tabcolsep{1.5pt}
\caption{Median values and 68\% confidence interval for the physical and orbital parameters of the KELT-21 system}
  \label{tbl:KELT-21b}
  \begin{tabular}{lccccc}
  \hline
  \hline
  Parameter & Description (Units) & \textbf{Adopted Value} & Value  \\
  & & \textbf{(YY circular)} & (Torres circular) \\
 \hline
Stellar Parameters & & & \\
                               ~~~$M_{*}$\dotfill &Mass (\msun)\dotfill & $1.458_{-0.028}^{+0.029}$&$1.526_{-0.067}^{+0.070}$\\
                             ~~~$R_{*}$\dotfill &Radius (\rsun)\dotfill & $1.638\pm0.034$&$1.663\pm0.041$\\
                         ~~~$L_{*}$\dotfill &Luminosity (\lsun)\dotfill & $8.03_{-0.53}^{+0.54}$&$8.28_{-0.58}^{+0.60}$\\
                             ~~~$\rho_*$\dotfill &Density (cgs)\dotfill & $0.468_{-0.024}^{+0.026}$&$0.468_{-0.024}^{+0.026}$\\
                  ~~~$\log{g_*}$\dotfill &Surface gravity (cgs)\dotfill & $4.173_{-0.014}^{+0.015}$&$4.180\pm0.016$\\
                  ~~~$\teff$\dotfill &Effective temperature (K)\dotfill & $7598_{-84}^{+81}$&$7600_{-84}^{+81}$\\
                                 ~~~$\feh$\dotfill &Metallicity\dotfill & $-0.405_{-0.033}^{+0.032}$&$-0.405\pm0.032$\\
             ~~~$v\sin{I_*}$\dotfill &Rotational velocity (\kms)\dotfill & $146.03\pm0.48$&$146.03_{-0.50}^{+0.49}$\\
           ~~~$\lambda$\dotfill &Spin-orbit alignment (degrees)\dotfill & $-5.6_{-1.9}^{+1.7}$&$-5.6_{-1.9}^{+1.7}$\\
         ~~~$NR Vel. W.$\dotfill &Non-rotating line width (\kms)\dotfill & $5.17\pm0.48$&$5.09\pm0.74$\\
         \hline
 Planet Parameters & & & \\
                                  ~~~$P$\dotfill &Period (days)\dotfill & $3.6127647\pm0.0000033$&$3.6127647\pm0.0000033$\\
                           ~~~$a$\dotfill &Semi-major axis (AU)\dotfill & $0.05224_{-0.00034}^{+0.00035}$&$0.05304_{-0.00079}^{+0.00080}$\\
                                 ~~~$M_{P}$\dotfill &Mass (\mj)\dotfill & $\color{red}(<3.91)$&$\color{red}(<4.07)$\\
                               ~~~$R_{P}$\dotfill &Radius (\rj)\dotfill & $1.586_{-0.040}^{+0.039}$&$1.610\pm0.045$\\
                           ~~~$\rho_{P}$\dotfill &Density (cgs)\dotfill & $\color{red}(<1.24)$&$\color{red}(<1.23)$\\
                      ~~~$\log{g_{P}}$\dotfill &Surface gravity\dotfill & $\color{red}(<3.59)$&$\color{red}(<3.59)$\\
               ~~~$T_{eq}$\dotfill &Equilibrium temperature (K)\dotfill & $2051_{-30}^{+29}$&$2051\pm29$\\
                           ~~~$\Theta$\dotfill &Safronov number\dotfill & $0.0048_{-0.0039}^{+0.025}$&$0.0050_{-0.0041}^{+0.026}$\\
                   ~~~$\fave$\dotfill &Incident flux (\fluxcgs)\dotfill & $4.01\pm0.23$&$4.02_{-0.22}^{+0.23}$\\
\hline
 Radial Velocity Parameters & & & \\
       ~~~$T_C$\dotfill &Time of inferior conjunction (\bjdtdb)\dotfill & $2457295.93434_{-0.00042}^{+0.00041}$&$2457295.93435\pm0.00042$\\
                        ~~~$K$\dotfill &RV semi-amplitude (m/s)\dotfill & $\color{red}(<399.6)$&$\color{red}(<406.3)$\\
                    ~~~$M_P\sin{i}$\dotfill &Minimum mass (\mj)\dotfill & $\color{red}(<3.91)$&$\color{red}(<4.07)$\\
                           ~~~$M_{P}/M_{*}$\dotfill &Mass ratio\dotfill & $\color{red}(<0.0025)$&$\color{red}(<0.0026)$\\
                       ~~~$u$\dotfill &RM linear limb darkening\dotfill & $0.5413_{-0.0033}^{+0.0058}$&$0.5412_{-0.0032}^{+0.0054}$\\
                            ~~~$\gamma_{McDonald}$\dotfill &\kms\dotfill & $-10.50\pm0.36$&$-10.51_{-0.36}^{+0.37}$\\
                                ~~~$\gamma_{TRES}$\dotfill &\kms\dotfill & $-10.72_{-0.18}^{+0.17}$&$-10.71\pm0.18$\\

\hline
\multicolumn{4}{l}{Linear Ephemeris from Follow-up Transits}            \\
                                 ~~~$P_{Trans}$\dotfill &Period (days)\dotfill & 3.6127628 $\pm$ 0.0000038 &---\\
       ~~~$T_0$\dotfill &Linear ephemeris from transits (\bjdtdb)\dotfill & 2457382.640727 $\pm$ 0.00041 &---\\
 \hline

 \hline
 \hline
 \end{tabular}
\begin{flushleft}
 \footnotesize \textbf{\textsc{NOTES}} \\
  \vspace{.1in}
  \footnotesize 3$\sigma$ limits are reported for KELT-21b's mass and parameters dependent on the planetary mass, and are shown in red. 
  \footnotesize The gamma velocity reported here uses an arbitrary zero point for the multi-order relative velocities. Uncertainties are based strictly on formal uncertaintes on the input data and do not take into account systematic effects. The quoted stellar parameters are for the planet host star KELT-21.
 \end{flushleft}
\end{table*}

\begin{table*}
\scriptsize
 \centering
\setlength\tabcolsep{1.5pt}
\caption{Median values and 68\% confidence intervals for the physical and orbital parameters for the KELT-21 System}
  \label{tbl:KELT-21b_part2}
  \begin{tabular}{lccccc}
  \hline
  \hline
  Parameter & Description (Units) & \textbf{Adopted Value} & Value \\
  & & \textbf{(YY circular)} & (Torres circular) \\
 \hline
 \hline
 Primary Transit & & & \\
~~~$R_{P}/R_{*}$\dotfill &Radius of the planet in stellar radii\dotfill & $0.09952_{-0.00073}^{+0.00071}$&$0.09951_{-0.00071}^{+0.00070}$\\
           ~~~$a/R_*$\dotfill &Semi-major axis in stellar radii\dotfill & $6.86_{-0.12}^{+0.13}$&$6.86_{-0.12}^{+0.13}$\\
                          ~~~$i$\dotfill &Inclination (degrees)\dotfill & $86.46_{-0.34}^{+0.38}$&$86.46_{-0.34}^{+0.38}$\\
                               ~~~$b$\dotfill &Impact parameter\dotfill & $0.423_{-0.039}^{+0.033}$&$0.423_{-0.039}^{+0.033}$\\
                             ~~~$\delta$\dotfill &Transit depth\dotfill & $0.00990\pm0.00014$&$0.00990\pm0.00014$\\
                    ~~~$T_{FWHM}$\dotfill &FWHM duration (days)\dotfill & $0.15242_{-0.00061}^{+0.00062}$&$0.15241_{-0.00063}^{+0.00062}$\\
              ~~~$\tau$\dotfill &Ingress/egress duration (days)\dotfill & $0.01864\pm0.00076$&$0.01863_{-0.00075}^{+0.00076}$\\
                     ~~~$T_{14}$\dotfill &Total duration (days)\dotfill & $0.17106_{-0.00092}^{+0.00091}$&$0.17104_{-0.00091}^{+0.00093}$\\
   ~~~$P_{T}$\dotfill &A priori non-grazing transit probability\dotfill & $0.1313_{-0.0023}^{+0.0022}$&$0.1313\pm0.0023$\\
             ~~~$P_{T,G}$\dotfill &A priori transit probability\dotfill & $0.1603_{-0.0030}^{+0.0028}$&$0.1603\pm0.0029$\\
                     ~~~$u_{1I}$\dotfill &Linear Limb-darkening\dotfill & $0.1486_{-0.0070}^{+0.012}$&$0.1487_{-0.0069}^{+0.011}$\\
                  ~~~$u_{2I}$\dotfill &Quadratic Limb-darkening\dotfill & $0.3067_{-0.012}^{+0.0099}$&$0.3065_{-0.012}^{+0.0098}$\\
                ~~~$u_{1Sloang}$\dotfill &Linear Limb-darkening\dotfill & $0.3475_{-0.0040}^{+0.0068}$&$0.3472_{-0.0039}^{+0.0063}$\\
             ~~~$u_{2Sloang}$\dotfill &Quadratic Limb-darkening\dotfill & $0.3442_{-0.0023}^{+0.0014}$&$0.3444_{-0.0021}^{+0.0013}$\\
                ~~~$u_{1Sloani}$\dotfill &Linear Limb-darkening\dotfill & $0.1662_{-0.0078}^{+0.013}$&$0.1663_{-0.0077}^{+0.012}$\\
             ~~~$u_{2Sloani}$\dotfill &Quadratic Limb-darkening\dotfill & $0.3123_{-0.012}^{+0.0100}$&$0.3122_{-0.012}^{+0.0099}$\\
                ~~~$u_{1Sloanr}$\dotfill &Linear Limb-darkening\dotfill & $0.2277_{-0.0076}^{+0.012}$&$0.2277_{-0.0076}^{+0.011}$\\
             ~~~$u_{2Sloanr}$\dotfill &Quadratic Limb-darkening\dotfill & $0.3370_{-0.0100}^{+0.0079}$&$0.3369_{-0.0095}^{+0.0079}$\\
                     ~~~$u_{1V}$\dotfill &Linear Limb-darkening\dotfill & $0.2901_{-0.0072}^{+0.011}$&$0.2900_{-0.0071}^{+0.011}$\\
                  ~~~$u_{2V}$\dotfill &Quadratic Limb-darkening\dotfill & $0.3365_{-0.0074}^{+0.0053}$&$0.3365_{-0.0070}^{+0.0053}$\\
                  \hline

Secondary Eclipse & & &\\
                  ~~~$T_{S}$\dotfill &Predicted time of eclipse (\bjdtdb)\dotfill & $2457294.12796_{-0.00042}^{+0.00041}$&$2457294.12796\pm0.00042$\\
\hline
\end{tabular}
\begin{flushleft}
 \footnotesize \textbf{\textsc{NOTES}} \\
  \vspace{.1in}
  \footnotesize Uncertainties are based strictly on formal uncertaintes on the input data and do not take into account systematic effects. 
 \end{flushleft}
\end{table*}

\subsection{Transit Timing Variation Analysis}
\label{sec:TTVs}

Using the fiducial global model determined transit center times (see Table \ref{tab:TTVs}), we searched for transit timing variations in the KELT-21 system. We ensure that all follow-up lightcurves are using BJD$_{\rm TDB}$ time stamps \citep{Eastman:2010}. Additionally, all follow-up members synchronize their telescope control computers to a standard clock prior to observing. This is typically done periodically throughout an observing night. We perform a linear fit to the determined transit mid times, obtaining a linear ephemeris of T$_0$ = 2457382.640727 $\pm$ 0.00041 (BJD$_{\rm TDB}$) and P = 3.6127628 $\pm$ 0.0000038 days, with $\chi^2$ = 27.2 and 6 degrees of freedom. Although several of the data points lie more than $1\sigma$ from the zero $O-C$ line, two of the greatest outliers are the observations from SUO (large scatter) and ZRO (ingress only) which we would expect to have lower timing precision. Furthermore, the WCO observations are of the same transit as that observed by KUO, which show no deviations from the linear ephemeris. These deviations are likely due to the systematics inherent to ground-based transit timing observations \citep{Carter:2009}. We thus find no conclusive evidence for any astrophysical TTVs in our data and adopt this as the best ephemeris for predicting future transit times. 

\begin{table}
\centering
 \caption{Transit times from KELT-21\MakeLowercase{b} Photometric Observations.}
 \label{tab:TTVs}
 \begin{tabular}{r@{\hspace{12pt}} l r r r c}
    \hline
    \hline
    \multicolumn{1}{c}{Epoch} & \multicolumn{1}{c}{$T_\textrm{C}$} 	& \multicolumn{1}{l}{$\sigma_{T_\textrm{C}}$} 	& \multicolumn{1}{c}{O-C} &  \multicolumn{1}{c}{O-C} 			& Telescope \\
	    & \multicolumn{1}{c}{(\bjdtdb)} 	& \multicolumn{1}{c}{(s)}			& \multicolumn{1}{c}{(s)} &  \multicolumn{1}{c}{($\sigma_{T_\textrm{C}}$)} 	& \\
    \hline
-134 & 2456898.527802  & 169 &  -233.74 &  -1.38 &    SUO\\
-118 & 2456956.337374  &  91 &   229.94 &   2.50 &    ZRO\\
  57 & 2457588.567597  &  67 &   -52.87 &  -0.79 &    CROW\\
  67 & 2457624.696694  &  69 &    74.02 &   1.07 &    KUO\\
  67 & 2457624.695694  &  72 &   -12.38 &  -0.17 &    KUO\\
  67 & 2457624.693194  &  60 &  -228.38 &  -3.80 &    WCO\\
 144 & 2457902.879452  &  67 &    75.71 &   1.13 &    ULMT\\
 144 & 2457902.879181  &  45 &    52.30 &   1.16 &    ULMT\\
    \hline
    \hline
 \end{tabular}
  \begin{flushleft}
  \footnotesize{Epochs are given in orbital periods relative to the value of the inferior conjunction time from the global fit.}
\end{flushleft}
\end{table} 

\begin{figure}

\includegraphics[width=1\linewidth,trim = 0.25in 0.2in 5.0in 9.0in]{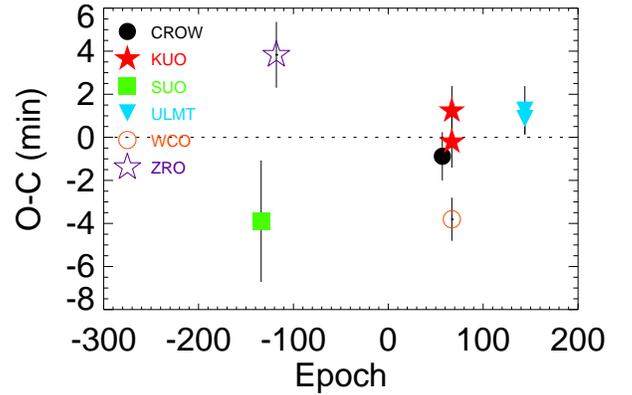}

\caption{The transit time residuals for KELT-21b using the inferior conjunction time from the global fit to define the epoch. The data are listed in Table \ref{tab:TTVs}.}
\label{fig:TTVs}
\end{figure}

\subsection{Tidal Evolution and Insolation History}
\label{sec:Insolation}

Using the POET code \citep{Penev:2014}, we followed the past and future tidal orbital evolution of the KELT-21 system under the constant phase lag (constant tidal quality factor) assumption. We incorporated the evolution of the stellar radius and luminosity, and followed the transfer of angular momentum from the star to the orbit. Note that in this case, the minimum equatorial velocity of the star (i.e., \vsinistar) implies that the star is spinning super-synchronously; given the stellar radius and \vsinistar\ found in \S\ref{sec:GlobalFit}, $P_{\mathrm{rot}}<0.57$ days, compared to $P=3.61$ days for the planet. Thus tidal dissipation causes the planet to move outward and the star to spin down. Calculations were performed with three different assumptions for the modified tidal quality factor of KELT-21: $Q_*' = 10^{6.03}$, $10^7$, and $10^8$ to demonstrate the range of plausible evolutions. The value of $Q_*' = 10^{6.03}$ was chosen as any smaller value of $Q_*'$ would require the planet to be inside the star at an age of 200 Myr. 
While tidal quality factors $>10^7$ are plausible, especially for ``hot'' stars like KELT-21, the orbital evolution on the main sequence is already small for $Q_*' = 10^7$, and differences in orbital evolution for $Q_*' = 10^7$ and $10^8$ are thus minimal.

\begin{figure}
\centering 
\includegraphics[width=1.0\columnwidth]{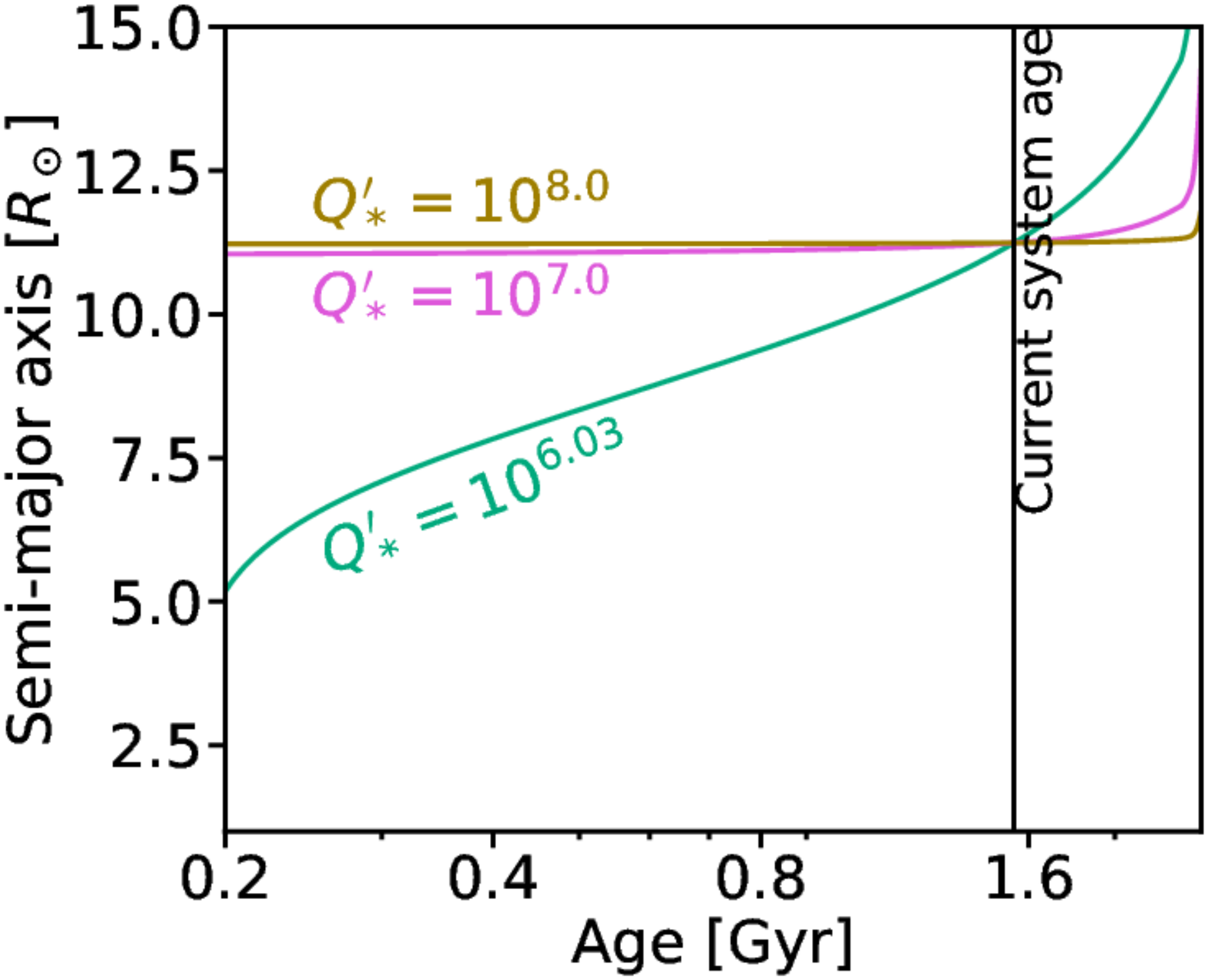}\\
\includegraphics[width=1.0\columnwidth]{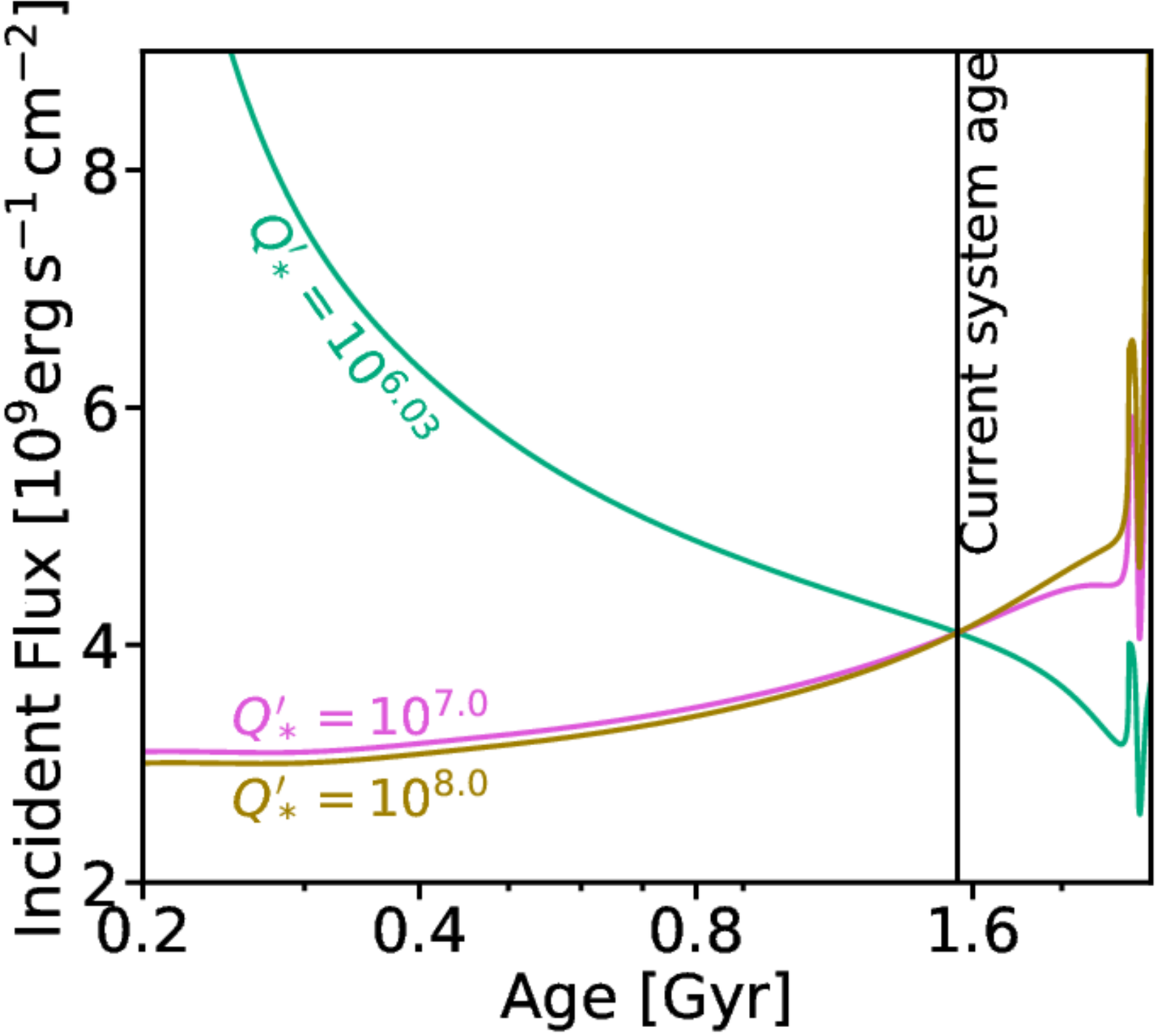}
\caption{\footnotesize Predicted evolution of the semi-major axis of (top) and stellar insolation experienced by (bottom) KELT-21b under the influence of stellar tides. Teal, magenta, and dark yellow lines depict the evolution assuming a stellar tidal quality factor of $Q_*' = 10^{6.03}$, $10^7$ and $10^8$, respectively.}
\label{fig:semimajoraxis}
\end{figure}

We show the evolution of the semi-major axis of KELT-21b in the top panel Fig.~\ref{fig:semimajoraxis}, and of the stellar insolation received by KELT-21b in the bottom panel. The changes in semi-major axis and insolation are small for $Q_*'\geq10^7$. For $Q_*'=10^{6.03}$ the evolution is more rapid and KELT-21b would have needed to begin its life at the stellar surface, and so the actual value of $Q_*'$ is likely to be larger than $10^{6.03}$.  Since we used the $3\sigma$ upper mass limit of $3.91$ \mj\ for these calculations, the value of $Q_*'$ could be smaller if the planetary mass is smaller, and the orbital evolution would be smaller for lower planetary mass at fixed $Q_*'$. 
Regardless of the value of $Q_*'$, larger changes will begin occurring in a few hundred million years, when KELT-21 begins evolving off of the main sequence.

\section{Discussion}
\label{sec:Discussion}

\subsection{Spin-Orbit Misalignment}
\label{sec:lambda}

With our Doppler tomographic observations we measured the spin-orbit misalignment, $\lambda$, obtaining $\lambda=-5.6_{-1.9}^{+1.7 \circ}$. This, however, is not the true obliquity of the planetary orbit, but rather this value projected onto the plane of the sky. The full three-dimensional spin-orbit misalignment $\psi$ is a more physically meaningful quantity than $\lambda$, but its calculation requires knowledge of not only $\lambda$ and the planetary orbital inclination $i$, but also the inclination of the stellar rotation axis $I_*$. The first two of these are straightforward to measure; the latter is not, and we cannot measure this quantity for KELT-21 using our existing data.

We can, however, set limits upon $\psi$ by making reasonable assumptions to limit $I_*$, as we did in \cite{Siverd:2017} and \cite{Lund:2017}. We follow \cite{Iorio:2011} by assuming that KELT-21 must be rotating at less than the break-up velocity, which must limit $I_*$, since the equatorial velocity is $v_{\mathrm{eq}}=\vsinistar/\sin I_*$. Doing so with our best-fit values for the system parameters, we obtain a break-up velocity of $v_{\mathrm{eq,max}}=232.5 \pm 3.3$ \kms, stellar inclination of $38.2^{\circ}<I_*<141.8^{\circ}$, and true orbital obliquity of $3.7^{\circ}<\psi<55.9^{\circ}$, the latter two ranges at $1\sigma$ confidence. While KELT-21b certainly has a prograde orbit, we cannot exclude the possibility that it is significantly misaligned with respect to the stellar rotation despite the alignment along the line of sight.

\begin{figure*}
\centering 
\includegraphics[scale=0.7]{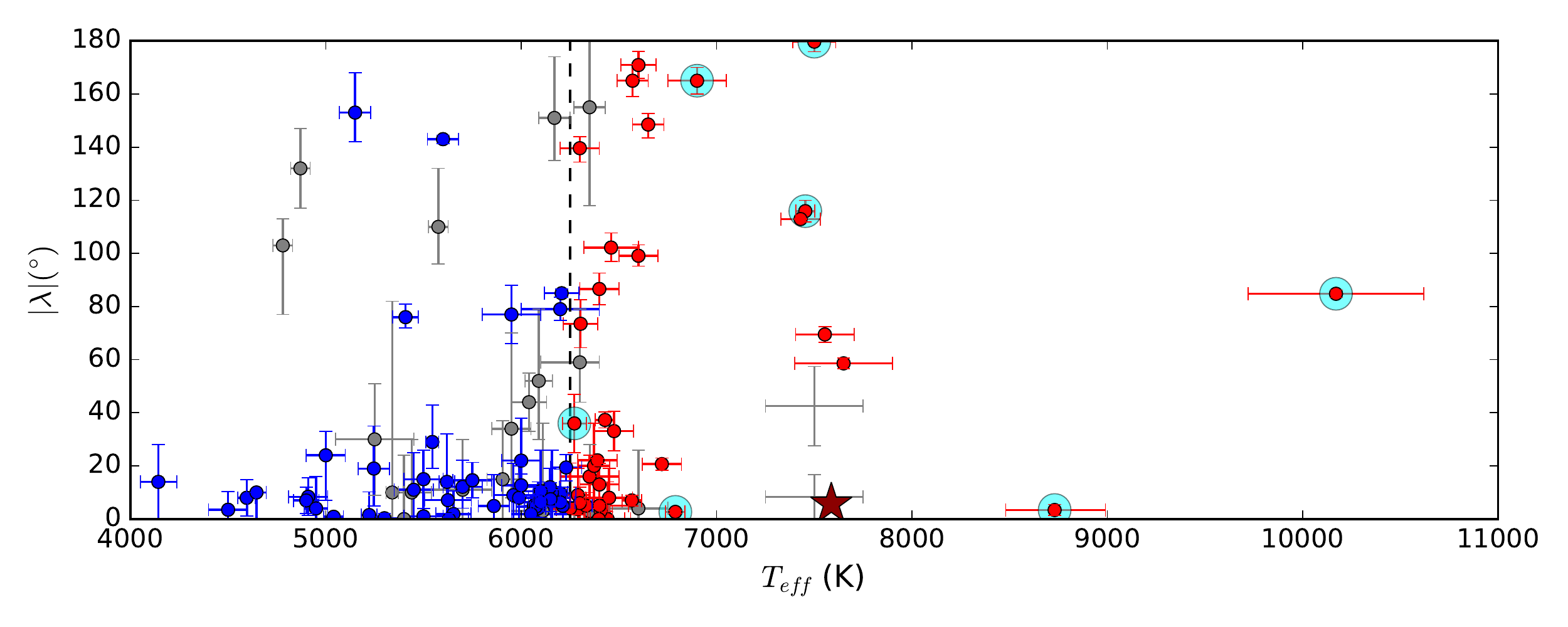}\\
\vspace{-10pt}
\includegraphics[scale=0.7]{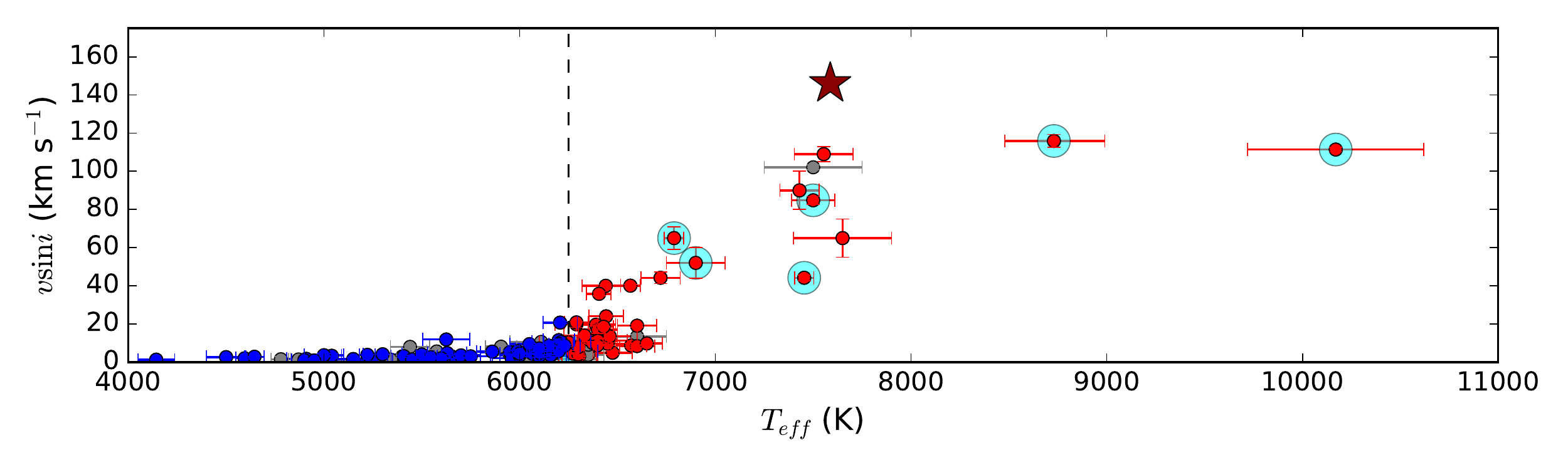}
\caption{\footnotesize Top: distribution of the spin-orbit misalignments of hot Jupiters from the literature as a function of $T_{\mathrm{eff}}$, after \cite{Winn:2010}. Planets with host stars with $T_{\mathrm{eff}}<6250$ K are shown in blue, those with $T_{\mathrm{eff}}>6250$ K in red, and those with uncertainties of $>20^{\circ}$ on their reported values of $\lambda$ in gray. A vertical dashed line denotes the approximate location of the Kraft break at 6250 K. Large cyan circles highlight planets discovered by the KELT survey, and KELT-21b is shown by the large dark red star; uncertainties on its parameters are smaller than the plot symbol size. HAT-P-57 is shown with the two sets of error bars without an associated plot point, denoting the two degenerate solutions found by \cite{Hartman:2015}. Bottom: $v\sin I_*$ values for these stars. The steady increase in average rotational velocity above the Kraft break is apparent. We assembled the literature sample using John Southworth's TEPCat Rossiter-McLaughlin Catalogue\protect\footnote{http://www.astro.keele.ac.uk/jkt/tepcat/}.}
\label{fig:lambdavsTeff}
\end{figure*}

In Fig.~\ref{fig:lambdavsTeff} we show the spin-orbit misalignments of all hot Jupiters for which this has been measured, along with KELT-21b. KELT-21 is well above the Kraft break, where hot Jupiters tend to have a wide range of misalignments \citep{Winn:2010}. Although its well-aligned orbit is somewhat unusual in this context, some other hot Jupiters around hot stars are also aligned, such as KELT-7b \citep{Zhou:2016} and KELT-20b/MASCARA-2b \citep{Lund:2017,Talens:2017}. Even for an isotropic distribution of misalignments, some systems should be aligned by chance. Indeed, it is perhaps unsurprising that KELT-21, as the most rapidly rotating hot Jupiter host star known to date (bottom panel of Fig.~\ref{fig:lambdavsTeff}), possesses an apparently well-aligned planet. In order for an aligned planet to transit, its host star must have $I_*\sim90^{\circ}$, and so on average the \vsinistar values for aligned planets should be higher than those for misaligned planets \citep[e.g.,][]{Schlaufman:2010,Winn:2017}.

The aligned orbit of KELT-21b suggests that it could have migrated to its current position through its protoplanetary disk, and that a dynamically hot migration mechanism, like planet-planet scattering \citep[e.g.,][]{Lin:1996} or the Kozai-Lidov mechanism \citep[e.g.,][]{Fabrycky:2007,Naoz:2012}, is not required to explain KELT-21b. The value of \feh$=-0.405_{-0.033}^{+0.032}$ that we have found is also interesting in the context of planet migration. \cite{Dawson:2013} found that hot and warm Jupiters orbiting lower-metallicity (\feh$<0$) stars typically have lower orbital eccentricities than those around higher-metallicity stars, and attributed this to disk migration for lower-metallicity stars and planet-planet scattering for higher-metallicity stars, which due to the planet-metallicity correlation \citep[e.g.,][]{Fischer&Valenti:2005} should form more giant planets. Planet-planet scattering should result in misaligned orbits, while disk migration should produce aligned orbits. That KELT-21, with \feh$<0$, has an aligned planet, could fit into this picture. On the other hand, the presence of possible stellar companions (\S\ref{sec:AO}) suggests that the Kozai-Lidov mechanism could still be responsible for the migration of KELT-21b; we discuss this possibility in more detail in \S\ref{sec:compimplications}.

\subsection{KELT-21 as a Hierarchical Triple System}
\label{sec:Triple}

\subsubsection{Are the Companions Bound?}

Candidate stellar companions discovered through high-resolution imaging are typically confirmed to be bound to the primary star through detection of common proper motion \citep[e.g.,][]{Ngo:2016}. Alternately, if only a single epoch of photometry is available, multi-color photometry can be used to check that the secondary is consistent with being at the same distance as the primary \citep[e.g.,][]{Evans:2016}. Finally, spectra can be used to determine whether the companion shares the systemic velocity of the primary \citep[e.g.,][]{Siverd:2017}. We, however, have only a single epoch of single-color photometry of the candidate companions found in \S\ref{sec:AO}, and the candidate companions are too faint to be detected in our spectra. We cannot apply any of these methods to confirm or assess whether the companions are bound.

Instead, we follow a similar methodology to \cite{Oberst:2016}, who used source counts from 2MASS to argue that the probability of a chance alignment between KELT-16 with a star at least as bright as their candidate companion was small. The companions to KELT-21, however, are more than two magnitudes fainter in $K_S$ than the companion to KELT-16, and are below the 2MASS completeness limit. We therefore used the same method as \cite{Oberst:2016}, but using star counts from a Galactic model rather than from actual data.

We generated Galactic models for the KELT-21 field using the v1.6 of the TRILEGAL code\footnote{http://stev.oapd.inaf.it/cgi-bin/trilegal\_1.6} \citep{Girardi:2005} and with the Besan\c{c}on code\footnote{http://model2016.obs-besancon.fr/} \citep{Robin:2003}. We generated a model for a one square degree field centered on the location of KELT-21 ($l=71.4814^{\circ}$, $b=-1.9865^{\circ}$). For TRILEGAL we used the extinction from the \cite{Schlafly:2011} reddening maps at this location\footnote{https://irsa.ipac.caltech.edu/applications/DUST/}; for the Besan\c{c}on model we used the dust map of \cite{Marshall:2006}. We assumed a binary fraction of 0.33 and flat mass ratio distribution between 0.2 and 1.0, after \cite{Raghavan:2010}, and otherwise used the default TRILEGAL and Besan\c{c}on parameters. We neglect the contribution of background galaxies as their source density is much smaller than that of stars near the Galactic plane. The source density of galaxies with $K<17$ is $\mathcal{O}(10^3)$ deg$^{-2}$ \citep[e.g.,][]{Smith:2009}, while both TRILEGAL and Besan\c{c}on predict a stellar source density of $\mathcal{O}(10^5)$ deg$^{-2}$ in the same magnitude range. We counted the number of model sources brighter than the combined $K_S$ magnitude of the companions ($K_S=16.47$), and approximated the probability of a chance superposition as this total times the fraction of the 1 square degree model area that is less than 1\farcs2 from KELT-21 (this being the separation of the actual companions). This resulted in a probability of 3.8\% (TRILEGAL) or 5.0\% (Besan\c{c}on) of a chance superposition. This suggests that the companions are likely to be bound, but the chance that they are background sources is too high to claim physical association with any certainty.

We can, however, further leverage the binary nature of the companion. First, we can assess the probability that the candidate companions are bound {\it to each other} using similar methodology. Here the probability of a chance superposition of an object brighter than $K_S=17.38$ within 55 mas of the brighter candidate companion is 0.013\% (TRILEGAL) or 0.019\% (Besan\c{c}on). We conclude that the companions are very likely to be bound to each other.

We can now assess the probability of the chance superposition of a {\it bound binary} close to KELT-21. To do so, we use the same TRILEGAL model as earlier, which accounts for the presence of binaries but does not contain any information on the separation of the binaries.  
For each binary in the TRILEGAL model, we drew a random orbital period from the log-normal distribution found by \cite{Raghavan:2010} to approximate their results (mean of $\log P=5.03$, standard deviation $\sigma_{\log P}=2.28$), converted this to semi-major axis using Kepler's Third Law and the masses of the components from TRILEGAL, and computed a projected separation using the simplifying assumption of circular orbits and drawing a random orbital phase. We then calculated the magnitude difference $\Delta K_S$ between the components using the \texttt{isochrones} code \citep{Morton:2015}, assuming the age, masses, and metallicity of the system from TRILEGAL. We then assessed the total number of binary systems with projected separations larger than our resolution limit of $\lambda/D=$45 mas, and smaller than the separation between KELT-21 and the candidate companions, and with a magnitude difference between the components smaller than that between the observed candidate companions. The Besan\c{c}on model does not automatically include binaries, and so we randomly assigned binary companions to 33\% of the model stars with a mass ratio drawn from a flat distribution over the interval $0.2\leq q \leq 1$ and then otherwise followed the same methodology as for TRILEGAL. This indicates a probability of 0.035\% (TRILEGAL) or 0.091\% (Besan\c{c}on) of the chance superposition of a background visual binary with KELT-21. We note that the choice of outer limit for this calculation has only a minimal effect on this result; 99\% of the binaries with appropriate contrast ratios have separations of less than 0\farcs5.

We thus conclude that the candidate companions are very likely to be bound to KELT-21, and for the remainder of this paper we assume that they are bound and refer to them as KELT-21B and KELT-21C. Future high-resolution imaging observations to confirm common proper motion of the companions will likely be difficult. The proper motion of KELT-21 is small (total proper motion of $2.34 \pm 0.80$ mas yr$^{-1}$; Table~\ref{tab:LitProps}), and, indeed, is non-zero at only $2.9\sigma$ in TGAS \citep{Brown:2016}. As is expected for a system close to the direction of Galactic rotation ($l=71^{\circ}$), the space motion of KELT-21 is primarily in the radial direction, resulting in small proper motion.

\subsubsection{Properties of the Companions}
\label{sec:compprop}

Assuming that the companions are indeed physically bound to KELT-21, we can estimate their physical properties.
In order to estimate the masses of the companions, we used the \texttt{isochrones} package \citep{Morton:2015} to calculate the expected $\Delta K_S$ values between a primary with the properties of KELT-21 and the companions as a function of companion mass, which we compared to the observed values. This resulted in estimated companion masses of $M_B=0.13_{-0.01}^{+0.02}$ \msun\ and $M_C=0.11 \pm 0.01$ \msun. The quoted uncertainties on these masses are derived purely from the uncertainties on the parameters of KELT-21 and on the photometric measurements, and neglect any systematic uncertainties from the isochrones or other sources. 
Using the $\mstar$-spectral type relations of \cite{Pecaut:2013}, these would correspond to spectral types of M5.5V and M6V for KELT-21 B and C, respectively.

Assuming $a\sim a_\perp$ \citep[which is true on average:][]{Heacox:1994}, the orbital periods of the B-C and A-BC binaries should be $\sim200$ and $\sim9000$ years, respectively. If the orbits were to be circular and face-on, this would correspond to an astrometric motion of A and BC due to their mutual orbit of $\sim120$ and $\sim750$ $\mu$as yr$^{-1}$, respectively, while the astrometric motion of B and C about each other would be $\sim800$ $\mu$as yr$^{-1}$.  
Gaia should be able to achieve a proper motion precision of $\sim3-8$ $\mu$as yr$^{-1}$ on KELT-21\footnote{https://www.cosmos.esa.int/web/gaia/science-performance}, and so should be able to easily detect its orbital motion in the A-BC binary. Components B and C should have $G$-band magnitudes of $\sim19.2$ \citep[using the relation between $G$, $V$, and $I_C$ magnitudes from][]{Jordi:2010}, and Gaia should be able to achieve a parallax precision of 130 $\mu$as and a proper motion precision of $\sim70$ $\mu$as yr$^{-1}$, sufficient to  both confirm that B and C are located at the same distance as KELT-21, and to detect the mutual orbital motion of the B-C and A-BC binaries. While the companions are undetected in the Gaia DR1 source catalog, this is not unexpected given the large flux difference and the incompleteness of this catalog within 4'' of bright sources \citep{Arenou:2017}. Given the low proper motion of KELT-21, it is likely that Gaia will confirm or refute the association of KELT-21 B and C with A before common proper motion confirmation is possible. 

\subsubsection{Implications of the Companions}
\label{sec:compimplications}

KELT-21b is likely one of the few known examples of a hot Jupiter in a hierarchical triple stellar system. Approximately 10\% of field systems with AFGK primaries are stellar triple systems \citep{Raghavan:2010,DeRosa:2014}. While theoretically it has been proposed that certain configurations of hierarchical triple systems should boost the efficiency of hot Jupiter formation \citep{Hamers:2017}, it is difficult to assess the occurrence rate of hot Jupiters in stellar triple systems due to observational biases against equal-mass triples and heterogenous imaging observations across the full sample of known hot Jupiters; doing so is beyond the scope of this paper.

For binary companions to exoplanet host stars, we can evaluate whether the companion is capable of causing Kozai-Lidov oscillations by equating the Kozai-Lidov and general relativistic precession timescales \citep[e.g.,][]{Ngo:2016}. Approximating KELT-21 BC as a single object with the sum of their masses, they would be capable of causing Kozai-Lidov oscillations of a giant planet with a semi-major axis of greater than $\sim2.1$ AU if the A-BC mutual orbit is circular; this limit decreases as the eccentricity of this orbit increases (e.g., to $\sim1.9$ AU for $e=0.5$). 

The dynamics of Kozai-Lidov oscillations due to a binary stellar companion, however, are more complicated than that due to a single companion \citep{Hamers:2017,Fang:2017}. \cite{Hamers:2017} found in their simulations that in hierarchical triple systems with a structure similar to that of KELT-21 (i.e., a primary with a planet and a pair of binary companions, with $a_{B-C} << a_{A-BC}$, which they referred to as a ``2+2'' configuration), hot Jupiters tended to be formed only if $a_{B-C}\lesssim10^2$ AU, and 20 AU $\lesssim a_{A-BC}\lesssim$ 10$^3$ AU. The period distribution of the resulting hot Jupiters in their models peaked around 3 days. Finally, \cite{Hamers:2017} found that Kozai-Lidov oscillations are enhanced if the Kozai-Lidov time scales for the planetary orbit, and the binary orbit of the companions, is similar; quantitatively, this occurs when $\mathcal{R}_{2+2}\sim(a_{B-C}/a_P)^{3/2}[($\mstar$+M_P)/(M_B+M_C)]^{3/2}\sim\mathcal{O}(1)$. For the KELT-21 system, if the planet formed at 5 (15) AU, the system would initially have had $\mathcal{R}_{2+2}\sim150$ ($30$). \cite{Hamers:2017} found that hot Jupiters tend to form in systems with $0.01\lesssim\mathcal{R}_{2+2}\lesssim100$. The KELT-21 system is broadly consistent with all of these criteria, suggesting that it is plausible that the companions drove the migration of the planet by four-body Kozai-Lidov oscillations.

Nonetheless, the well-aligned orbit of KELT-21 is somewhat at odds with the Kozai-Lidov migration scenario. Hot Jupiters formed by this mechanism should frequently reside on highly-inclined orbits \citep[cf. Fig.~9 of][]{Hamers:2017}. On the other hand, we found in \S\ref{sec:lambda} that KELT-21b has $\psi<54.7^{\circ}$, which is reasonable given the distribution of $\psi$ (there denoted $\theta_*$) found by \cite{Hamers:2017}. 

We note in conclusion that, given the many migration mechanisms that have been proposed to create hot Jupiters, it is always difficult or impossible to ascribe the formation of a specific system to a specific mechanism with any certainty. Instead, it is the distributions of parameters of a population (e.g., $P$, $\lambda$, etc.), that constrains the migration mechanisms. KELT-21b adds to the population of known hot Jupiters around hot stars, and thus will help to answer these questions statistically.

\subsection{Metal Content and Galactic Context}
\label{sec:metals}

\begin{figure*}
\includegraphics[width=1.0\columnwidth]{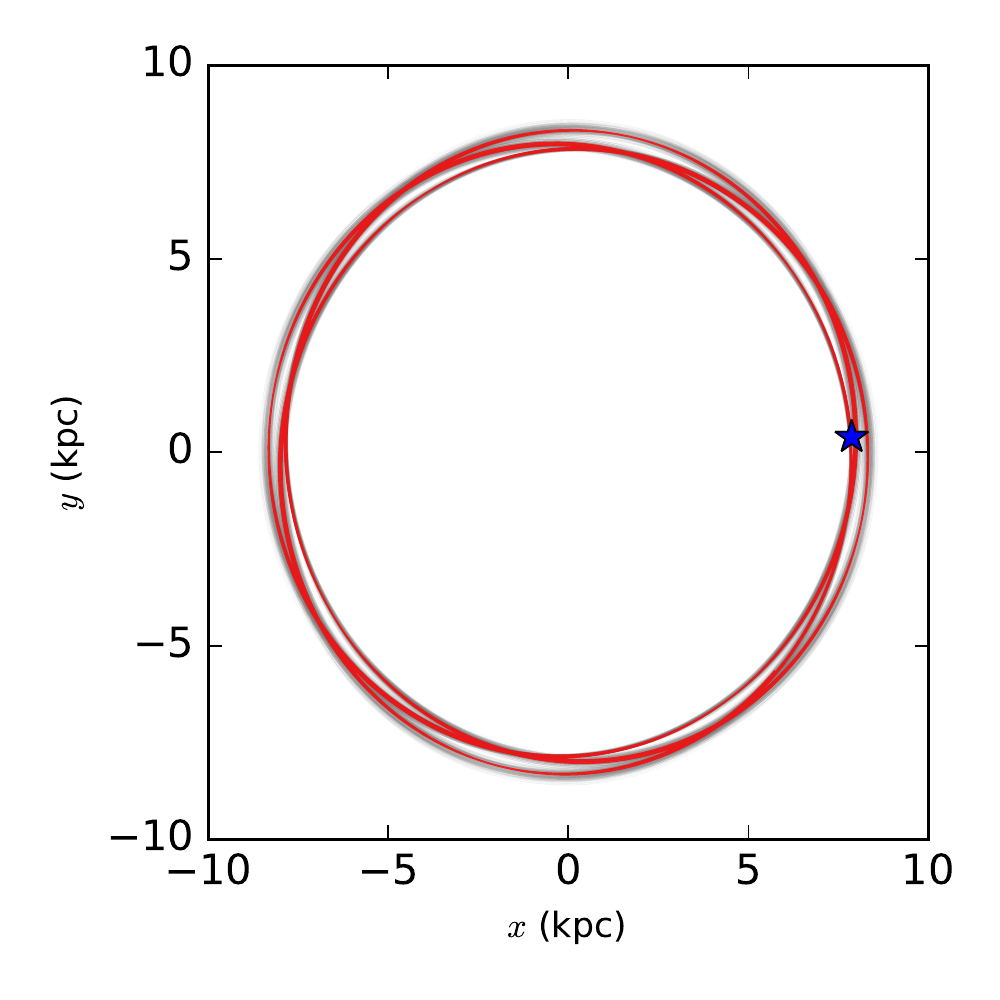}
\includegraphics[width=1.0\columnwidth]{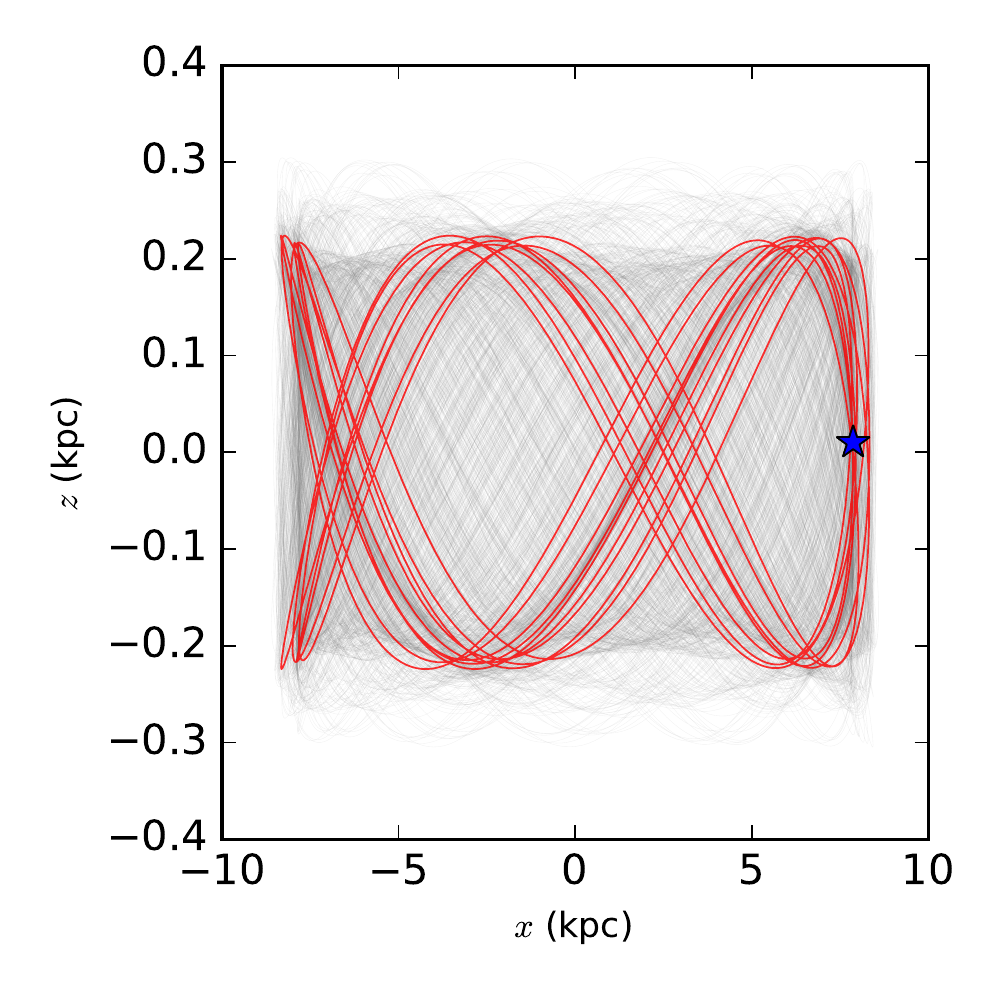}
\caption{Galactic orbit of KELT-21, as computed with \texttt{galpy} from KELT-21's position, distance, and ($U,V,W$) space velocity, as viewed from the top in the $x-y$ plane (left) and the side in the $x-z$ plane (right). The red line shows the orbit computed over the next 2 Gyr using our computed parameters of KELT-21, while the gray lines show 50 realizations with values of ($U,V,W$) and distance randomly drawn from Gaussian distributions with the same mean and standard deviation as the measured value and uncertainty on each parameter. The blue star marks the current position of KELT-21. KELT-21 does not stray far from the solar circle or the Galactic plane, confirming its thin disk kinematics.}
\label{fig:galorbit}
\end{figure*}

The $\alpha$ enhancement and relatively low metallicity found by our spectral analysis of KELT-21 (\S\ref{sec:SpecPars}) are unusual for a relatively young ($\sim1.6$ Gyr) 
star. In order to better contextualize the low metallicity, we computed the Galactic orbit of KELT-21 using the \texttt{galpy} package\footnote{http://github.com/jobovy/galpy} \citep{Bovy:2015}. Given our ($U$,$V$,$W$) values and using \texttt{galpy}'s ``MWPotential2014'' Galactic potential, we estimate that KELT-21's Galactic orbit has an apoapsis of at most $\sim8.6$ kpc and a periapsis of at minimum $\sim7.7$ kpc. This suggests that KELT-21 does not stray far from the solar circle, but its low metallicity could still be explained by invoking formation in the metal-poor outer Galaxy or Galactic bar followed by radial mixing due to the Milky Way's spiral arms \citep[e.g.,][]{Sellwood:2002}. See Fig.~\ref{fig:galorbit} for a depiction of KELT-21's Galactic orbit. KELT-21 does not attain a height above the Galactic midplane of more than $\sim300$ pc, indicating that it is kinematically a member of the thin disk and substantiating our calculation that it has a high probability of being a member of the thin disk (\S\ref{sec:UVW}). Its [$\alpha$/Fe] value, however, is more similar to thick disk than thin disk stars at its \feh\ \citep[cf.][]{Bovy:2016,SilvaAguirre:2017}, although it does lie in between the thin and thick disk chemical sequences.

\begin{figure}
\includegraphics[width=1.1\columnwidth]{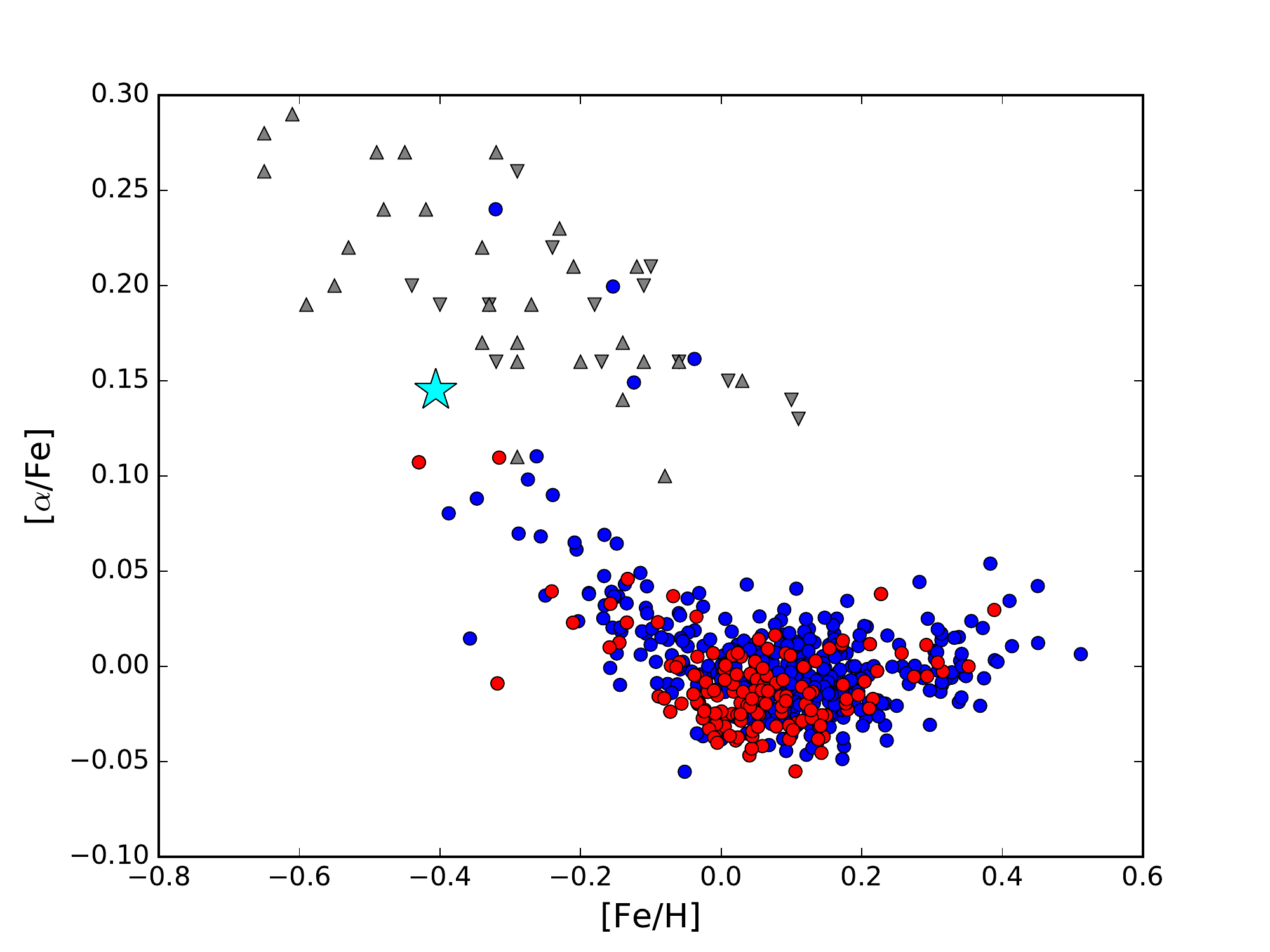}
\caption{$\alpha$-enhancement as a function of \feh\ for the APOKASC samples of $1.75 \msun<\mstar<2.25 \msun$ giants (blue) and relatively young giants with kinematics similar to KELT-21 (red; see text for details). KELT-21 is shown as the large cyan star. We also show the young $\alpha$-rich stars found by \cite{Martig:2015} and \cite{Chiappini:2015} as downward- and upward-pointing gray triangles, respectively. Note that \cite{Martig:2015} quote [Fe/M] and [$\alpha$/M], not \feh\ and [$\alpha$/Fe], and so the placement of these points on our plot should be considered approximate.}
\label{fig:fehvsalpfe}
\end{figure}

In order to assess how unusual KELT-21 is for a relatively young field star, we utilized the APOKASC sample of stars with abundances, masses, and ages \citep[Pinsonneault et al. in prep., an update to the][sample]{Pinsonneault:2014}. These stars have asteroseismic parameters measured from \textit{Kepler} photometry \citep{Borucki:2010}, masses inferred from asteroseismic scaling relations \citep{Kjeldsen:1995} with theoretical corrections \citep{Serenelli:2017} and empirical calibrations to cluster data. They also have spectroscopic metallicities, temperatures, and abundances from Data Release 14 \citep{DR14} of the APOGEE-2 survey \citep{Majewski:2017,Holtzman:2015} of the Sloan Digital Sky Survey IV \citep{Blanton:2017} on the Sloan Digital Sky Survey telescope \citep{Gunn:2006}. For our comparison, we assessed the values of \feh\ and [$\alpha$/Fe] for two subsamples: giants with masses of $1.75 \msun<\mstar<2.25 \msun$, which should have formed at around the same time as KELT-21, and stars with ages of less than 1.7 Gyr and Gaia parallaxes from which we could compute Galactic kinematics, again using \texttt{galpy} \citep{Bovy:2015}. For the latter sample we selected stars with Galactic periapses $>6.6$ kpc, and apoapses $<10.0$ kpc, in order to produce a sample with similar kinematics to KELT-21. For both samples we excluded stars with [C/N]$>-0.4$ in order to exclude stars which are likely merger products or have experienced significant mass gain due to binary interactions, and are therefore likely older than would be assumed given their other properties \citep[cf.][]{Izzard:2017}. We show the resulting \feh\ and [$\alpha$/Fe] values in Fig.~\ref{fig:fehvsalpfe}. It is apparent that KELT-21's \feh\ is at the lower end of these distributions, but it is not a dramatic outlier. 
We thus conclude that while KELT-21's low metallicity is unusual for a relatively young star that is kinematically part of the thin disk, it is not inexplicably so.

KELT-21's abundance pattern and relatively young age is qualitatively similar to those of the class of young, $\alpha$-rich giants found through the combination of {\it Kepler}/{\it CoRoT} asteroseismology and APOGEE spectra by \cite{Martig:2015} and \cite{Chiappini:2015}, although KELT-21 falls at the edge of the distribution of such stars (Fig.~\ref{fig:fehvsalpfe}). KELT-21 could be such a star observed while it is still on the main sequence. Such stars found by \cite{Chiappini:2015} were primarily in the inner Galactic disk, and they invoked formation at the end of the Galactic bar; while this is inconsistent with KELT-21's current circular orbit and location near the solar circle, as mentioned earlier it could still have formed in the inner Galaxy and experienced radial mixing to move it to its current location, although this would have needed to be rapid in order to move the star several kpc within the 1.6 Gyr since it formed. The planet orbiting KELT-21 also seems to be at odds with the other proposed explanation for young $\alpha$-rich stars, namely, that they are blue stragglers formed from stellar mergers \citep{Jofre:2016,Yong:2016}. Such a collision would have destroyed any short-period planet already around one of the stars; KELT-21b would have needed to either form from material thrown off in the collision, or have survived the collision at larger semi-major axis and only migrated {\it after} the collision. Young $\alpha$-rich stars also tend to be kinematically members of the thick disk, while KELT-21 is not, and so the association of KELT-21 with this class is not clear.

KELT-21's low metallicity is also unusual for a hot Jupiter host. Giant planets are much more common around more metal-rich stars; this is the well-known planet-metallicity correlation \citep[e.g.,][]{Fischer&Valenti:2005}. Only a handful of hot Jupiters are known to orbit stars more metal-poor than KELT-21, the current record-holder being WASP-98, with \feh=-0.60 \citep{Hellier:2014}. KELT-21b is thus useful for probing the properties of giant planets at low metallicity. Additionally, planet-host stars at low metallicity tend to be significantly $\alpha$-enhanced \citep[e.g.,][]{Adibekyan:2012}.  
It is thought that this is because stars with higher [$\alpha$/Fe] have more metals available to form solids (and therefore planets) at fixed \feh. 
KELT-21's $\alpha$-enhancement is in between the values typical for the thick and thin disk populations, which is not inconsistent with this trend.

\subsection{Prospects for Characterization}
\label{Characterization}

\begin{figure}
\centering 
\includegraphics[width=1.1\columnwidth]{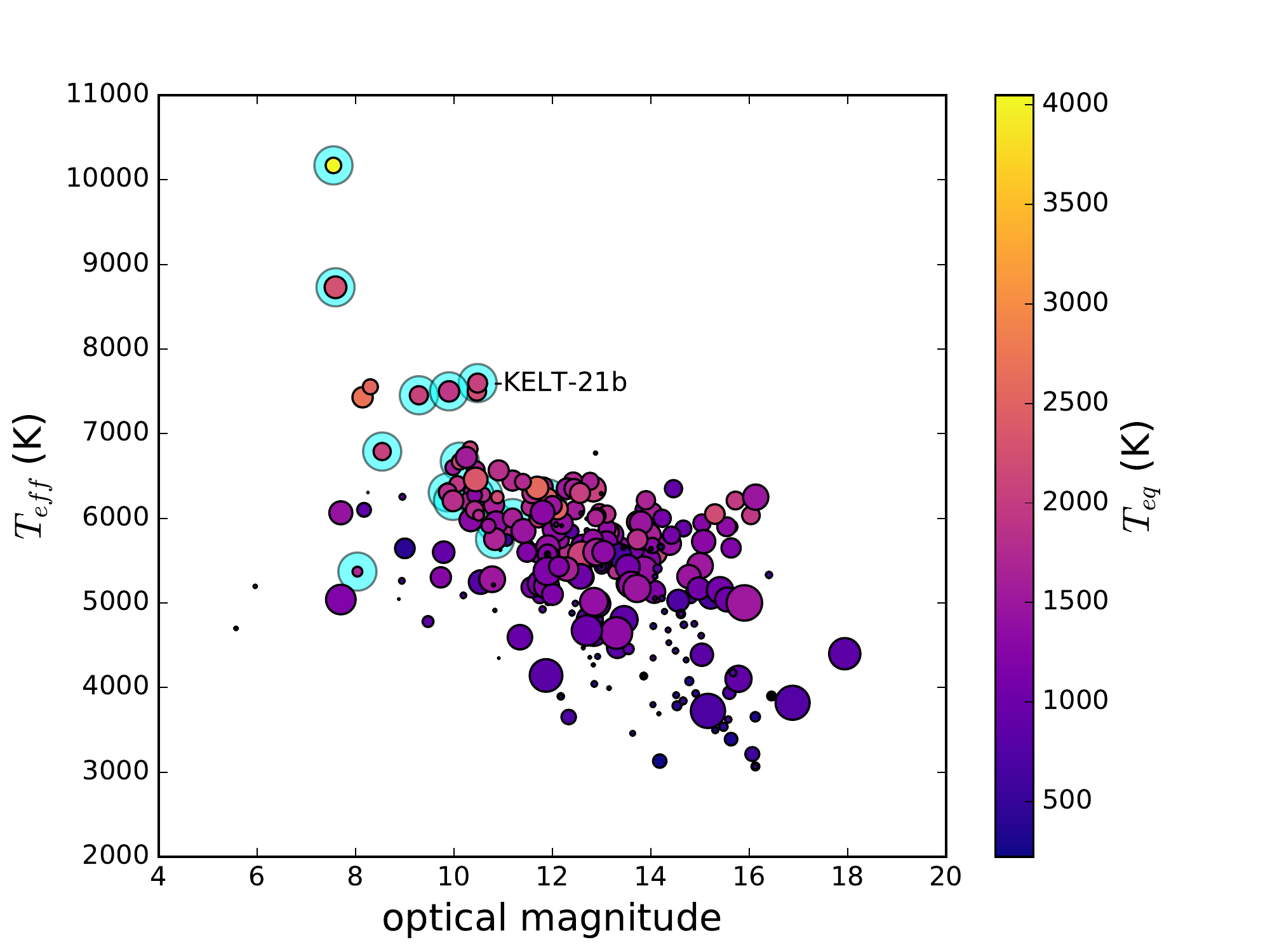}
\caption{\footnotesize KELT-21 in context with other known planets for the purpose of atmospheric investigations. The symbol color corresponds to the planetary zero-albedo equilibrium temperature, while the symbol size is proportional to the transit depth $(R_P/R_*)^2$. Planets discovered by KELT are marked with large cyan circles. KELT-21 is hotter and brighter than most transiting exoplanet hosts, and prospects for observations to characterize the atmosphere are good. The optical magnitude on the horizontal axis is the $Kp$ magnitude for {\it Kepler} planets and $V$ magnitude for most other objects. We show known transiting planets taken from the NASA Exoplanet Archive\footnote{https://exoplanetarchive.ipac.caltech.edu/}, plus KELT-19Ab \citep{Siverd:2017}, KELT-20b/MASCARA-2b \citep{Lund:2017,Talens:2017}, and MASCARA-1b \citep{Talens:2017b}. The location of KELT-21b is marked.}
\label{fig:popplot}
\end{figure}

Although KELT-21 is, at $V=10.5$, one of the fainter transiting planet hosts found by KELT, it is nonetheless brighter than many transiting planet hosts (cf. the {\it Kepler} sample), and so prospects for further characterization are good. In Fig.~\ref{fig:popplot} we show KELT-21 in context of the population of known transiting planets, in terms of host star optical magnitude and \teff. Only a few stars hotter than KELT-21 host transiting hot Jupiters, and due to its relatively long-period orbit compared to these planets (3.6 days) KELT-21b is relatively cool (zero-albedo equilibrium temperature of $T_{\mathrm{eq}}=2051$ K), potentially offering interesting prospects for atmospheric characterization via transmission spectroscopy.  

KELT-21 is also relatively bright in the infrared ($J=10.15$, $K_S=10.09$; Table~\ref{tab:LitProps}), suggesting that KELT-21b may also be a good target for secondary eclipse observations in the infrared. Either alone or in concert with transmission spectroscopy during the transit, this will help to constrain the atmospheric properties of KELT-21b.

Due to the high \vsinistar\ of KELT-21, we have been unable to measure the mass of KELT-21b, only to set a $3\sigma$ upper limit of $M_P<3.91$ \mj. Future observations might be able to measure the mass; however, the rapid rotation of KELT-21  will make this very difficult. Another possible avenue to measure the mass and probe the atmosphere of KELT-21b is through the detection of the orbital phase curve \citep[e.g.,][]{Shporer:2011}. We estimate that, in the TESS bandpass, KELT-21 should have an orbital phase curve amplitude of $\sim65$ ppm. KELT-21 (TIC 203189770) has a TESS-bandpass magnitude of 10.33, which should result in a per-point photometric precision of $\sim240$ ppm \citep[using the noise-$T$ magnitude relationship presented in][]{Stassun:2017}, and will be on TESS silicon for up to 54.8 days in TESS Year 2 (as KELT-21 is in the northern ecliptic hemisphere). This precision and duration of data, folded over the orbital period of the planet and binned, results in a precision of $\sim60$ ppm, suggesting that the phase curve of KELT-21b should be detectable by TESS. Due to the high equilibrium temperature, the TESS-bandpass phase curve should be dominated by thermal emission from the planet (amplitude $\sim35$ ppm), which would be even more prominent in the infrared; this may also be detectable with {\it Spitzer}. 

\section{Summary and Conclusion}

We have presented the discovery of KELT-21b, a hot Jupiter on a 3.6-day orbit transiting the rapidly rotating A8V star HD 332124. 
With \vsinistar=146 \kms, KELT-21 is the most rapidly rotating star to host a confirmed transiting planet to date, and KELT-21b is one of only a handful of known transiting planets around an A star. Its host star is also relatively bright, suggesting good prospects for follow-up observations to further characterize the planet. 

Our high-resolution imaging observations revealed the presence of a close pair of faint stars at a separation of 1\farcs2 from the planet host star. Although we cannot confirm using our current data whether they are physically associated with the KELT-21 system, we have argued statistically that they are unlikely to be background sources. If they are indeed physically associated with KELT-21, KELT-21 B and C are a pair of mid-M dwarfs with a mutual separation of $\sim20$ AU, lying $\sim500$ AU from KELT-21. They occupy the part of parameter space where they could have caused the migration of KELT-21b through the Kozai-Lidov mechanism \citep[e.g.,][]{Hamers:2017}, although the well-aligned orbit of KELT-21b is not entirely consistent with such an origin.

Unusually for a star of its relatively high mass and thus relatively young age, KELT-21 appears to have a somewhat low metallicity (\feh=$-0.405_{-0.033}^{+0.032}$) and an $\alpha$ enhancement ([$\alpha$/Fe]=$0.145 \pm 0.053$). While this metallicity is unusual for a relatively young ($\sim1.6$ Gyr) star with thin-disk kinematics, it is not inexplicably so, and the [$\alpha$/Fe] is more typical of thick-disk stars at this \feh. KELT-21b is also among the lowest-metallicity stars known to host a hot Jupiter, and is thus particularly interesting in the context of planet formation theory.

\section{Acknowledgements}

We thank Jennifer Johnson, Marc Pinsonneault, and Dennis Stello for useful discussions on the Galactic context of KELT-21 and the analysis using the APOKASC catalog.

Work performed by J.E.R. was supported by the Harvard Future Faculty Leaders Postdoctoral fellowship.
Work by G.Z. is provided by NASA through Hubble Fellowship grant HST-HF2-51402.001-A awarded by the Space Telescope Science Institute, which is operated by the Association of Universities for Research in Astronomy, Inc., for NASA, under contract NAS 5-26555.
Work performed by P.A.C. was supported by NASA grant NNX13AI46G.
K.P. acknowledges support from NASA grant NNX13AQ62G.
B.S.G. and D.J.S. were partially supported by NSF CAREER Grant AST-1056524.
A.S. is partially supported by grant ESP2015-66134-R.
Funding for the Stellar Astrophysics Centre is provided by The Danish National Research Foundation (Grant agreement No. DNRF106). V.S.A. acknowledges support from VILLUM FONDEN (research grant 10118).

This project makes use of data from the KELT survey,
including support from The Ohio State University, Vanderbilt
University, and Lehigh University, along with the KELT
follow-up collaboration. The LBT is an international collaboration among institutions in the United States, Italy and Germany. The LBT Corporation partners are: The Ohio State University; LBT Beteiligungsgesellschaft, Germany, representing the Max Planck Society, the Astrophysical Institute Potsdam, and Heidelberg University; The University of Arizona on behalf of the Arizona university system; Istituto Nazionale di Astrofisica, Italy; The Research Corporation, on behalf of The University of Notre Dame, University of Minnesota and University of Virginia. This paper includes data taken at The McDonald Observatory of The University of Texas at Austin. Some of the data presented herein were obtained at the W.M. Keck Observatory, which is operated as a scientific partnership among the California Institute of Technology, the University of California and the National Aeronautics and Space Administration. The Observatory was made possible by the generous financial support of the W.M. Keck Foundation. The authors wish to recognize and acknowledge the very significant cultural role and reverence that the summit of Mauna Kea has always had within the indigenous Hawaiian community.  We are most fortunate to have the opportunity to conduct observations from this mountain. This work has made use of NASA's Astrophysics Data System, the Extrasolar Planet Encyclopedia at exoplanet.eu, the SIMBAD database operated at CDS, Strasbourg, France, and the VizieR catalogue access tool, CDS, Strasbourg, France.  We also used data products from the Wide-field Infrared Survey Explorer, which is a joint project of the University of California, Los Angeles, and the Jet Propulsion Laboratory/California Institute of Technology, funded by the National Aeronautics and Space Administration; the Two Micron All Sky Survey, which is a joint project of the University of Massachusetts and the Infrared Processing and Analysis Center/California Institute of Technology, funded by the National Aeronautics and Space Administration and the National Science Foundation; and the European Space Agency (ESA) mission {\it Gaia} (\url{http://www.cosmos.esa.int/gaia}), processed by the {\it Gaia} Data Processing and Analysis Consortium (DPAC, \url{http://www.cosmos.esa.int/web/gaia/dpac/consortium}). Funding for the DPAC has been provided by national institutions, in particular the institutions participating in the {\it Gaia} Multilateral Agreement.

\facilities{KELT, LBT(PEPSI), FLWO(TRES), HJST(TS23), Keck(NIRC2)}
\software{Python, IDL, IRAF, TAPIR \citep{Jensen:2013}, AstroImageJ \citep{Collins:2013}, SDS4PEPSI \citep{Strassmeier:2017}, EXOFAST \citep{Eastman:2013}, POET \citep{Penev:2014}, TRILEGAL \citep{Girardi:2005}, Besan\c{c}on \citep{Robin:2003}, isochrones \citep{Morton:2015}, galpy \citep{Bovy:2015}}

\bibliography{KELT-21b}{}
\bibliographystyle{apj}

\end{document}